\documentclass{emulateapj}

\def\tobs{t_{\rm obs}}

\def\gch{\gamma_{\rm ch}}

\def\nuobs{\nu_{\rm obs}}

\def\alphaB{\alpha_{\rm B}}
\def\betaB{\beta_{\rm B}}

\def\be{\begin{equation}}
\def\ee{\end{equation}}
\def\beq{\begin{eqnarray}}
\def\eeq{\end{eqnarray}}

\begin{document}

\title{Toward an understanding of GRB prompt emission mechanism: \\
II. Patterns of peak energy evolution and their connection to spectral lags}

\author{Z. Lucas Uhm\altaffilmark{1,2,\dag}, Bing Zhang\altaffilmark{2}, and Judith Racusin\altaffilmark{1}}

\altaffiltext{1}{Astrophysics Science Division, NASA Goddard Space Flight Center, Greenbelt, MD 20771, USA}
\altaffiltext{2}{Department of Physics and Astronomy, University of Nevada, Las Vegas, NV 89154, USA}
\altaffiltext{\dag}{NASA Postdoctoral Program (NPP) Senior Fellow; z.lucas.uhm@gmail.com}

\begin{abstract}
The prompt emission phase of gamma-ray bursts (GRBs) exhibits two distinct patterns of the peak-energy ($E_p$) evolution; i.e., time-resolved spectral analyses of $\nu F_{\nu}$ spectra of broad pulses reveal (1) ``hard-to-soft'' and (2) ``flux-tracking'' patterns of $E_p$ evolution in time, the physical origin of which still remains not well understood. We show here that these two patterns can be successfully reproduced within a simple physical model invoking synchrotron radiation in a bulk-accelerating emission region. We show further that the evolution patterns of the peak energy have, in fact, direct connections to the existence of two different (positive or negative) types of spectral lags, seen in the broad pulses. In particular, we predict that (1) only the positive type of spectral lags is possible for the hard-to-soft evolution of the peak energy, (2) both the positive and negative type of spectral lags can occur in the case of flux-tracking pattern of the peak energy, (3) for the flux-tracking pattern, the peak location of the flux light curve slightly lags behind the peak of the $E_p$ evolution with time if the spectral lags are positive, and (4) in the case of flux-tracking pattern, double-peaked broad pulses can appear in the light curves, the shape of which is energy-dependent.
\end{abstract}

\keywords{gamma-ray burst: general --- radiation mechanisms: non-thermal --- relativistic processes}

%--------------------------------------------------------
%
% Section 1
%
%--------------------------------------------------------

\section{Introduction}

Although it is agreed that the gamma-ray bursts (GRBs), the most energetic electromagnetic explosions in the universe, invoke highly relativistic jets with bulk Lorentz factors of a few hundreds, the exact physical mechanism producing such powerful gamma-rays still remains debated \citep[e.g.,][for a recent review]{kumarzhang15}. Three outstanding questions in the field concern (1) the composition of GRB jets, (2) the involved radiative process responsible for the observed gamma-rays, and (3) the distance of the emitting region from the central engine where the prompt gamma-rays are released.

One class of proposed models invokes a matter-dominated outflow. \cite{paczynski86} and \cite{goodman86} considered an optically-thick ``fireball'' made of electron-positron plasma and photons, which gives rise to thermal blackbody radiation from the fireball photosphere at a photospheric radius $\sim 10^{11} - 10^{12}$ cm. \cite{shemi90} examined the influence of baryonic matter on the fireball expansion. An optically-thin region above the photosphere was then considered where the internal shocks resulting from the relativistic unsteady outflow emit non-thermal (synchrotron and/or inverse Compton) radiation at a typical internal-collision radius $\sim 10^{13} - 10^{14}$ cm \citep{rees94,daigne98}. Motivated by steep low-energy spectral slopes observed in some GRBs, \cite{meszarosrees00} examined the role of a photospheric component and Comptonization in the internal shock model. \cite{rees05} introduced a dissipative photospheric model where an additional energy dissipation occurs below the baryonic photosphere, suggesting that the GRB spectral peak is essentially due to the Comptonized thermal component of the photosphere.

An alternative class of proposed models invokes a Poynting-flux-dominated outflow. \cite{thompson94} considered a relativistic, strongly-magnetized outflow, which generates the gamma-ray emission via inverse Compton of seed photons at a small distance. He also considered a magnetically-dissipative photosphere picture where a thin layer of Wolf-Rayet material entrained by the jet head becomes transparent \citep{thompson06}. \cite{drenkhahn02} showed that local magnetic-energy dissipation in a Poynting-flux-powered outflow efficiently accelerates the flow to a high bulk Lorentz factor through magnetic-pressure gradient both below and above the photosphere, and for typical GRB parameters the dissipation takes place mainly above the photosphere, producing non-thermal radiation up to a saturation radius of $\sim 10^{13} - 10^{14}$ cm. \cite{mckinney12} used the magnetohydrodynamical models of ultra-relativistic jets and invoked a switch from the slow collisional to fast collisionless reconnection regime, to produce GRBs at a radius of $\sim 10^{14}$ cm.

Motivated by solving several issues of GRB prompt emission, such as the missing photosphere problem, low efficiency problem, low electron number problem, and inconsistency between prompt emission correlations with the internal shock model, \cite{zhangyan11} proposed the internal-collision-induced magnetic reconnection and turbulence (ICMART) model. This model envisages that the GRB central engine launches an intermittent, magnetically-dominated outflow, where fast reconnection and relativistic turbulence induced by internally-colliding mini-shells, similar to the internal-shock model, result in a runway release of the stored magnetic energy at a relatively large distance $\sim 10^{15} - 10^{16}$ cm, producing synchrotron radiation to power the observed gamma-rays. This model was further developed with a different trigger mechanism due to the kink instability \citep{lazarian18}. 

During the prompt emission phase of GRBs, the observed gamma-ray spectra typically show a smoothly-connected broken power-law shape and are usually well described by a phenomenological ``Band-function'' \citep{band93}. In recent years, extensive efforts have been made in modeling this shape of prompt emission spectra \citep[e.g.,][]{meszarosrees00,peer06,beloborodov10,lazzati10,vurm11,daigne11,lundman13,uhm14}, and some detailed direct comparison to the observational data have also been made in different contexts \citep[e.g.,][]{burgess14,ahlgren15,zhangbb16}.

Needless to say, it is clear that any viable physical model for GRB prompt emission needs to interpret both the spectral and temporal behavior of the observed emission. The prompt gamma-ray light curves display diverse and complex features such that the light curves of thousands of observed GRBs are all essentially different from one another. Nevertheless, a large fraction of those complicated light curves contains an interesting, common characteristic, i.e., the existence of a single or multiple ``broad pulses'' \citep{norris96,hakkila11} that are either separated from or overlapped with one another and that are slowly varying as opposed to the rapid variabilities.

Noteworthily, two important properties of broad pulses are observationally revealed. The first of those regards two distinct patterns of the peak-energy ($E_p$) evolution across the broad pulses \citep[e.g.,][]{hakkila11,lu12}. The time-resolved spectral analyses of $\nu F_{\nu}$ spectra of broad pulses reveal either ``hard-to-soft'' \citep{norris86} or ``flux-tracking'' \citep{golenetskii83,kargatis94,bhat94} patterns of $E_p$ evolution in time. The second important property is the so-called ``spectral lags'' between the pulse light curves at different energies \citep[e.g.,][]{cheng95,norris96,band97,norris00,wu00,liang06}. Namely, a broad pulse's light curves at different frequencies exhibit a sequential pattern in their peak time with systematic time lags or spectral lags between those light curves. These two features are usually connected \citep{kocevski03,ryde00,borgonovo01}. In most cases, the observed spectral lags are ``positive''; i.e., the higher the energy of a light curve, the earlier the peak time of the light curve. A small fraction of the observed pulses shows an opposite pattern, i.e., ``negative'' spectral lags (where the higher the energy of a light curve, the later the peak time of the light curve) or no spectral lags.

The physical origin of these rich observational features still remains not-properly-understood. As it is clear that these distinct patterns displayed by the broad pulses carry important observational clues to unveil the physical mechanism of GRB prompt emission, we started to systematically model these features in a series of papers. In the first paper \citep{uhm16b}, we studied the origin of spectral lags and showed that the traditional view invoking high-latitude emission ``curvature effect'' \citep[e.g.,][]{dermer04,shen05} cannot account for the spectral lags. Instead, we showed that the observed spectral lags are successfully reproduced within a simple physical model that invokes synchrotron radiation emitted from a bulk-accelerating outflow at a large distance ($\sim 10^{15}-10^{16}$ cm) from the central engine.

In this second paper, in addition to modeling the spectral lags, we also produce the two distinctive patterns of the peak-energy ($E_p$) evolution across the broad pulses and show that the $E_p$ evolution pattern has, in fact, close and direct connections to the occurrence of positive or negative spectral lags, with some predicted properties that can be tested against observations in the future. We briefly summarize our physical picture in Section~\ref{section:2} and present the results of our numerical models in Section~\ref{section:3}. In Section~\ref{section:4}, we conclude the paper with Conclusions and Discussion.

%--------------------------------------------------------
%
% Section 2
%
%--------------------------------------------------------

\section{A simple physical model} \label{section:2}

In this paper, we adopt the same physical picture as in the first paper \citep{uhm16b}, in which a thin relativistic spherical shell expands in space radially and emits photons uniformly from all locations in the shell. In the co-moving frame of the shell, the emission produced at every location has an isotropic angular distribution for the emitted power, and the shape of the emission spectrum is described by a functional form \citep{uhm15},
\be
\label{eq:Hx}
H(x)
\quad \mbox{with} \quad
x=\nu^{\prime}/\nu_{\rm ch}^{\prime}.
\ee
This concept of giving a shape of the photon spectrum without specifying a specific radiative process allows the physical picture to remain general and is particularly useful in dealing with the relativistic effects between the co-moving frame and the observer frame \citep{uhm15}. The function $H(x)$ here has an arbitrary shape and is a function of the frequency $\nu^{\prime}$. A characteristic frequency $\nu_{\rm ch}^{\prime}$ indicates a characteristic location of the spectrum in the frequency space. Both frequencies $\nu^{\prime}$ and $\nu_{\rm ch}^{\prime}$ are measured in the co-moving frame. A uniform radiation power from all locations in the shell is given by a uniform distribution of radiating electrons that are placed in the shell. We assume that the total number of the radiating electrons $N$ in the shell increases at an injection rate, $R_{\rm inj} \equiv dN/dt^{\prime}$, from an initial value of $N = 0$. The time $t^{\prime}$ here is measured in the co-moving frame. Lastly, the electrons in the shell are assumed to have a same value of spectral power $P_0^{\prime}$ (measured in the co-moving frame), thus ensuring the uniformity of the shell's radiation. We remark that the two quantities $\nu_{\rm ch}^{\prime}$ and $P_0^{\prime}$ of characterizing the emission do not necessarily remain at a constant value as the shell propagates in space.

The GRB explosions occur at cosmological distances from the Earth. In the local lab frame of a GRB, the prompt emission is produced at a certain distance from the explosion center, and thus we consider that the emission is turned on at a radius $r_{\rm on}$ and at a lab-frame time $t_{\rm on}$. Upon the receipt of first photons from this turn-on point along the line of sight, an observer on the Earth sets an observer time $\tobs$ be equal to zero. Subsequent photons emitted at a radius $r$ $(>r_{\rm on})$ and at a lab-frame time $t$ $(>t_{\rm on})$ are then detected by the observer at observer time \citep{uhm16b}
\be
\label{eq:tobs}
\tobs=\left[ \left(t-\frac{r}{c} \cos\theta \right)-\left(t_{\rm on}-\frac{r_{\rm on}}{c}\right) \right] (1+z).
\ee
Here, $c$ is the speed of light, and $z$ is the cosmological redshift of GRB site. The photons are emitted from a spherical shell, and thus the angle $\theta$ here denotes for the latitude of their emission location measured from the observer's line of sight. As the shell travels radially with a profile of the Lorentz factor $\Gamma(r)$, the lab-frame time $t$ can be calculated as $t = t_{\rm on} + \int_{r_{\rm on}} dr/(c\beta)$ starting from the turn-on point, where $c\beta$ is the speed of the shell as given by $\beta = (1 - 1/\Gamma^2)^{1/2}$.

We follow the formulation given in \cite{uhm15} and take fully into account the high-latitude emission effect of the spherical shell, by including the relativistic Doppler boosting from the shell co-moving frame to the lab frame for each latitude and by considering the delayed arrival time of emitted photons for each emission latitude as given by Equation (\ref{eq:tobs}). For each observer time $\tobs$, we integrate over its equal-arrival-time surface (EATS) and find the observed spectral flux, $F_{\nu_{\rm obs}}^{\,\rm obs}$, as a function of two variables $\tobs$ and $\nuobs$. Here, $\nuobs$ is the observed frequency of photons (when detected by the observer) and is given by
\be
\nuobs = \nu^{\prime}\, [\Gamma (1-\beta \cos\theta)]^{-1}\, (1+z)^{-1}.
\ee
The angle $\theta$ here denotes for the latitude of emission location again.

In our first paper \citep{uhm16b}, we showed that the spectral lags and their observed properties can be successfully modeled within this simple physical picture while invoking synchrotron radiation, provided that (1) the emission spectrum $H(x)$ is curved, (2) the strength of magnetic field $B(r)$ in the emitting region globally decreases with radius as the emitting shell travels, and (3) the emitting region itself undergoes rapid bulk acceleration (i.e., an increasing profile of $\Gamma(r)$) during which the prompt gamma-rays are released. The observed gamma-ray spectra are indeed curved and are usually well described by the Band function \citep{band93}. The second requirement is naturally expected for a jet expanding in a 3-dimensional space and was the essential physical ingredient to explain the low-energy photon index of the Band function for a majority of prompt emission spectra \citep{uhm14}. The third requirement of bulk acceleration is recently evidenced by an independent analysis as well, made on the steep decay phase of GRB X-ray flares \citep{uhm16a,jia16}.\footnote{It is suggested that the steep decay phase of GRB X-ray flares may be partially interpreted with anisotropic synchrotron radiation invoked in the co-moving frame of the jet \citep{beloborodov11b,geng17}, but in order to fully reproduce the data, the bulk acceleration of the jet is still required.}

Hence, we follow these findings here. For the functional form $H(x)$ of giving the emission spectrum in the co-moving frame, we take a Band-function shape as
\beq
\label{eq:Hx}
H(x)=
\left\{
\begin{array}{ll}
x^{\alphaB+1} \exp(-x)                \quad & \mbox{if} \quad x \leq x_c, \\
(x_c)^{x_c} \exp(-x_c)\, x^{\betaB+1} \quad & \mbox{if} \quad x \geq x_c,
\end{array}
\right.
\eeq 
where $x_c \equiv \alphaB-\betaB$. Note that the indices $\alphaB$ and $\betaB$ are the low- and high-energy photon index of this smoothly-connected two power-law shape, respectively. Synchrotron radiation is then invoked to give the characteristic frequency $\nu_{\rm ch}^{\prime}$ and the spectral power $P_0^{\prime}$ of the electrons as follows \citep{rybicki79}
\be
\label{eq:syn}
\nu_{\rm ch}^{\prime}=\frac{3}{16}\, \frac{q_{\rm e} B}{m_{\rm e} c}\, \gch^2, 
\quad
P_0^{\prime} = \frac{3\sqrt{3}}{32}\, \frac{m_{\rm e} c^2\, \sigma_{\rm T} B}{q_{\rm e}},
\ee
where $q_{\rm e}$ and $m_{\rm e}$ are the electron charge and mass, respectively, and $\sigma_{\rm T}$ is the Thomson cross section. The strength of magnetic fields $B$ and the characteristic Lorentz factor $\gch$ of the electrons in the shell are measured in the co-moving frame.

The redshift affects the observed spectral flux in a global manner \citep{uhm15}, and thus we take a typical value of $z=1$ in all numerical models presented in this paper. The luminosity distance to the GRB explosion is calculated for a flat $\Lambda$CDM universe with the cosmological parameters $\Omega_{\rm m} = 0.31$, $\Omega_{\rm \Lambda} = 0.69$, and $H_0 = 68$ km $\mbox{s}^{-1}$ $\mbox{Mpc}^{-1}$ \citep{Planck16}.

%--------------------------------------------------------
%
% Section 3
%
%--------------------------------------------------------

\section{Results of example models} \label{section:3}

We begin with the three numerical models presented in \cite{uhm16b}, which are named as [2b], [2c], and [2d]. These three models have the following parameters. The low- and high-energy photon index of the Band-function shape $H(x)$ is $\alphaB=-0.8$ and $\betaB=-2.3$, respectively. The number of radiating electrons in the shell increases at a constant injection rate $R_{\rm inj}=10^{47}$ $\mbox{s}^{-1}$. The Lorentz factor profile $\Gamma (r)$ of showing an accelerating bulk motion of the shell takes a power-law form in radius as follows
\be
\label{eq:bulk}
\Gamma(r)=\Gamma_0 (r/r_0)^{s},
\ee
where a normalization value $\Gamma_0$ is set to be $\Gamma_0=250$ at radius $r_0=10^{15}$ cm with an index $s=0.35$. The index $s$ here describes a degree of bulk acceleration. The emission of the spherical shell is turned on at a turn-on radius $r_{\rm on}=10^{14}$ cm, and we turn off its emission at a turn-off radius $r_{\rm off}=3 \times 10^{16}$ cm. For the bulk motion given in Equation (\ref{eq:bulk}), this turn-off radius corresponds to a turn-off time at about $\tobs=4.0$ s. The strength of magnetic fields $B(r)$ in the co-moving frame has also a power-law profile in radius
\be
\label{eq:B}
B(r) = B_0 (r/r_0)^{-b},
\ee
with a normalization value $B_0=30$ G at radius $r_0=10^{15}$ cm. The index $b$ is set to be 1.0, 1.25, and 1.5 for the models [2b], [2c], and [2d], respectively.\footnote{One can consider a simple flux-conservation of magnetic fields that are frozen in a spherical jet expanding in a 3 dimensional space and get $b=1$ for the toroidal component. Also, the strength of magnetic fields can decrease faster than the case of $b=1$ due to possible dissipation of magnetic energy via the reconnection of field lines.} Lastly, the characteristic Lorentz factor $\gch$ of the electrons in the shell takes $\gch=5 \times 10^4$ for all three models [2b], [2c], and [2d]. Hence, the model parameters of these three models differ only by the $b$ index. Note that these parameters were adjusted to assure that the observed duration of broad pulses in the prompt gamma-ray light curves is about a few seconds and the observed peak energy $E_p$ of $\nu F_{\nu}$ spectra is of the order of 1 MeV.

Beginning with these three models, we present a total of twenty numerical models. As in the models [2b], [2c], and [2d], the letters `b', `c', and `d' contained in a model name will always indicate for a decreasing strength of magnetic fields, given in Equation (\ref{eq:B}), with the $b$ index 1.0, 1.25, and 1.5, respectively. Also, the number `2' in the beginning of any model names is to indicate that the emitting region of those models undergoes bulk acceleration as shown in Equation (\ref{eq:bulk}). One single index $s=0.35$ will be used in all twenty models.

\begin{figure}
\begin{center}
\includegraphics[width=18cm]{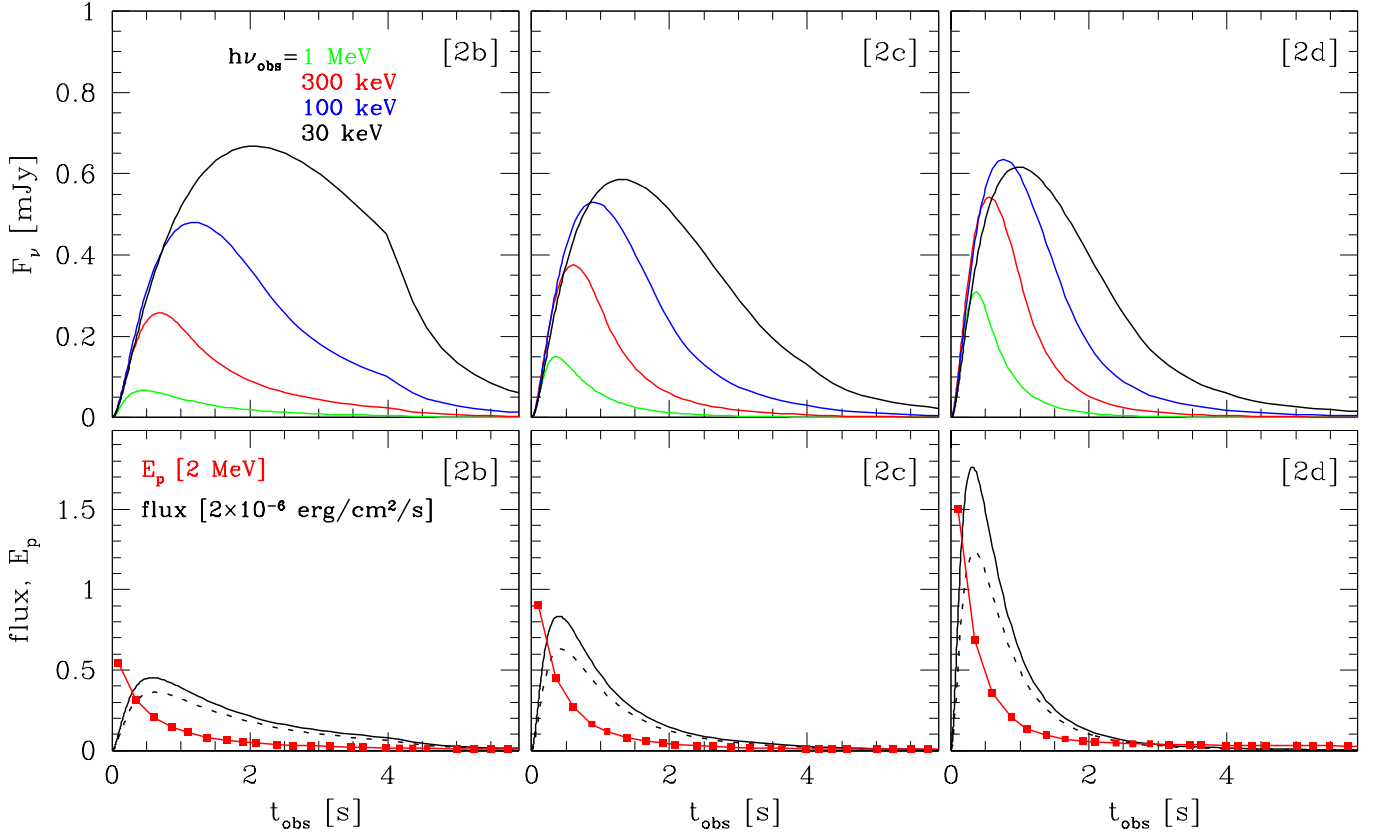}
\caption{
Observed spectral flux, total flux, and peak energy, emitted from a relativistically expanding spherical shell, for the numerical models [2b], [2c], and [2d]. Top panels show the four different light curves at 30 keV (black), 100 keV (blue), 300 keV (red), and 1 MeV (green). Bottom panels show the temporal curves for the total flux (solid-black), a detector-energy-range flux from 10 keV to 10 MeV (dashed-black), and the peak energy $E_p$ of $\nu F_{\nu}$ spectra (red). Left, middle, and right column corresponds to the model [2b], [2c], and [2d], respectively. We consider a globally-decreasing strength of magnetic fields, $B(r) \propto r^{-b}$, in the co-moving frame of the shell, and the index $b$ is set to be 1.0, 1.25, and 1.5 for the models [2b], [2c], and [2d], respectively. For detailed physical picture and the model parameters, see the text. In particular, the characteristic Lorentz factor $\gch$ of the electrons in the shell takes a constant value here in all three models.
}
\label{fig:f1}
\end{center}
\end{figure}

The calculation results for the three models [2b], [2c], and [2d] are shown in Figure~\ref{fig:f1}. The top panels show the four different light curves at 30 keV (black), 100 keV (blue), 300 keV (red), and 1 MeV (green). The bottom panels show the temporal curves for the peak energy $E_p$ (red) and the observed flux (solid-black). The dashed black curves in the bottom panels show the flux received in a detector-energy-range from 10 keV to 10 MeV. The left, middle, and right column corresponds to the model [2b], [2c], and [2d], respectively, as indicated by the model name shown in the upper right corner in each panel. An abrupt decrease at about $\tobs=4$ s in the low-energy light curves of the model [2b] is caused by our sudden turning-off of the shell's emission at the turn-off radius $r_{\rm off}$, and thus is not likely to be physical. As one can see, in all three models, the light curves exhibit a clear pattern of positive spectral lags while the $E_p$ temporal curve exhibits a hard-to-soft evolution. We point out here that a decreasing profile of $B(r)$ with $b \geq 1$ provides a natural ground for the hard-to-soft evolution of $E_p$ since the frequency $\nuobs$ along the observer's line of sight roughly follows $\nuobs \propto \Gamma B \propto r^{s-b}$ with $s-b < 0$.

\begin{figure*} \centering
\centering
\begin{tabular}{cc}
\includegraphics[width=8.27cm]{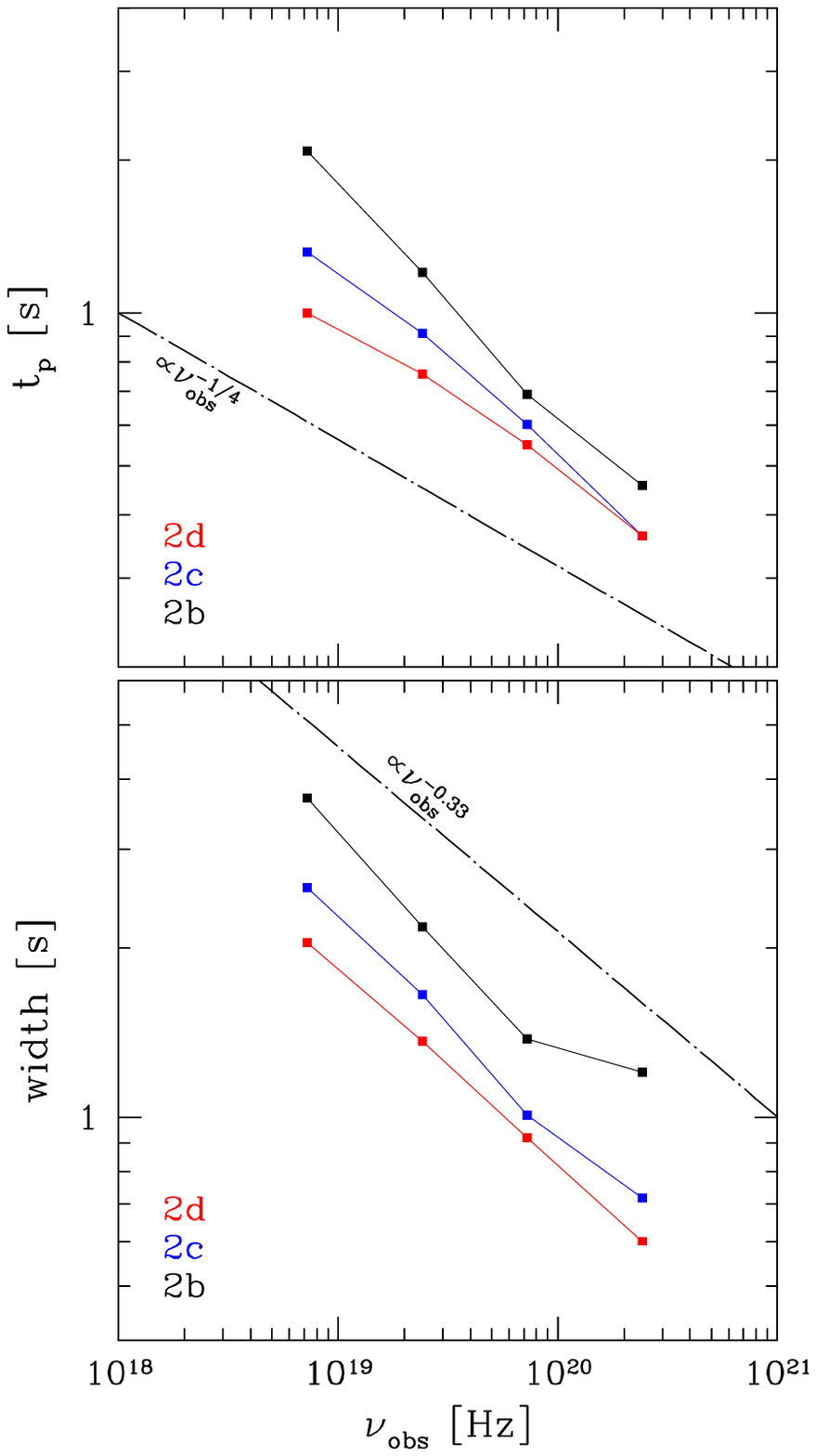} &
\includegraphics[width=8.50cm]{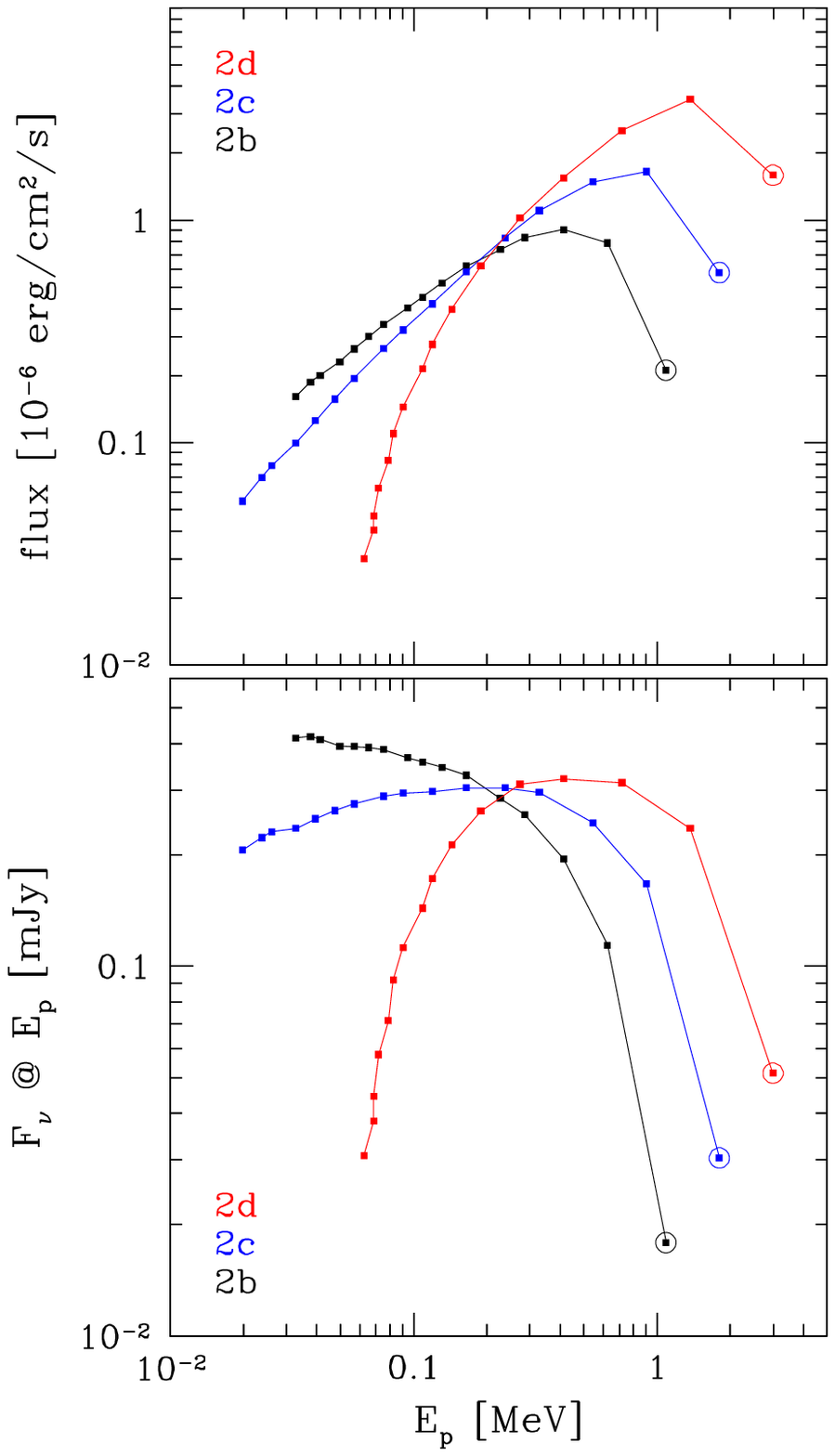} \\
\end{tabular}
\caption{
Properties of broad pulses and the pattern curves of the peak evolution, for the numerical models [2b], [2c], and [2d]. Left column shows the peak time $t_p$ (top panel) and the width (bottom panel) of four broad-pulse light curves in each model, which are also compared to the observations indicated by the dot-dashed lines \citep{norris96,liang06}. Note that these two dot-dashed lines here are meant to show the slope only and thus are plotted with an arbitrary normalization. Right column shows a time-evolving pattern curve of the ($E_p$, flux) points (top panel) and the ($E_p$, $F_{\nu, E_p}$) points (bottom panel) in each model. The point enclosed by an open circle in each model marks the first in observer time among the plotted points.
}
\label{fig:f2}
\end{figure*}

Figure~\ref{fig:f2} shows some detailed properties of the models [2b], [2c], and [2d]. In the upper left panel, we show for each of the models the four different points ($\nuobs$, $t_p$) and connect them by a solid line where $\nuobs$ and $t_p$ are, respectively, the frequency and the peak time of each of the four different light curves in the model. Hence, a negative slope in this panel indicates the positive type of spectral lags. In the lower left panel, we repeat the same, but instead of showing the peak time $t_p$, we show the width of the light curves where the width of a broad pulse is calculated as the full width at the half maximum of the pulse. The obtained curves are also compared to the dot-dashed lines of showing the relations $t_p \propto \nuobs^{-1/4}$ and $width \propto \nuobs^{-0.33}$, which are revealed by observations \citep{norris96,liang06}. Note that the two dot-dashed lines here are meant to show the slope only and thus are plotted with an arbitrary normalization. The upper right panel shows the flux against the peak energy $E_p$ for each model while the lower right panel shows the peak spectral flux $F_{\nu, E_p}$ against $E_p$ for the three models. Here, $F_{\nu, E_p}$ is the observed spectral flux measured at the location of $E_p$, namely, $F_{\nu}\, @\, E_p$. The points shown in these two panels are with $\tobs < 4$ s only, so as to avoid any effects caused by the sudden turning-off of the shell's emission. The point enclosed by an open circle in each model marks the point with the earliest observer-time $\tobs$ among the plotted points in the model. Therefore, a ``counter-clockwise'' evolving pattern is evident in all three models [2b], [2c], and [2d], which show a hard-to-soft pattern of $E_p$ evolution with the positive type of spectral lags in Figure~\ref{fig:f1}. As we will demonstrate with more examples below, this counter-clockwise pattern will be a ``defining signature'' of the positive type of spectral lags. We also point out that the two panels in the right column are closely related to each other since the flux is very roughly given by $(E_p/h)\, F_{\nu, E_p}$. Here, $h$ is the planck constant. Therefore, the panel with ($E_p$, flux) points contains an underlying linear relationship by default. After this linear relationship being removed, the panel with ($E_p$, $F_{\nu, E_p}$) points displays more informative pattern of the peak evolution, which will become clear with more examples below. As we are aware that a figure like the upper right panel is more often presented in the literature, we in this paper will stress on the usefulness of the lower right panel.

Now in attempts of reproducing various patterns of the peak evolution, including the $E_p$ tracking-behavior with the flux, we explore one very intuitively clear method, in which the characteristic Lorentz factor $\gch$ of electrons in the shell is allowed to evolve as the shell propagates in space. Initially, we consider a single power-law profile in radius
\be
\label{eq:gch}
\gch(r) = \gch^0 (r/r_0)^g,
\ee
where a normalization value $\gch^0$ is set to be $\gch^0 = 5 \times 10^4$ at radius $r_0 = 10^{15}$ cm with an index $g = -0.2$. Besides this profile of $\gch(r)$, we keep all other model parameters the same as in the models [2b], [2c], and [2d] and name three new models as [2b$_i$], [2c$_i$], and [2d$_i$], respectively. The subscript $i$ here indicates for this $\gch$ profile, which is shown in Figure~\ref{fig:f3}. Note that Figure~\ref{fig:f3} also shows the $\gch$ profiles of other example models to be presented below; in the paper, we explore five different variations on the $\gch$ profile, which are indicated by five subscripts $i$, $j$, $k$, $l$, and $m$ contained in the model names. Also, see Table~\ref{table:parameters} that summarizes the model parameters of our numerical models.

\begin{figure}
\begin{center}
\includegraphics[width=18cm]{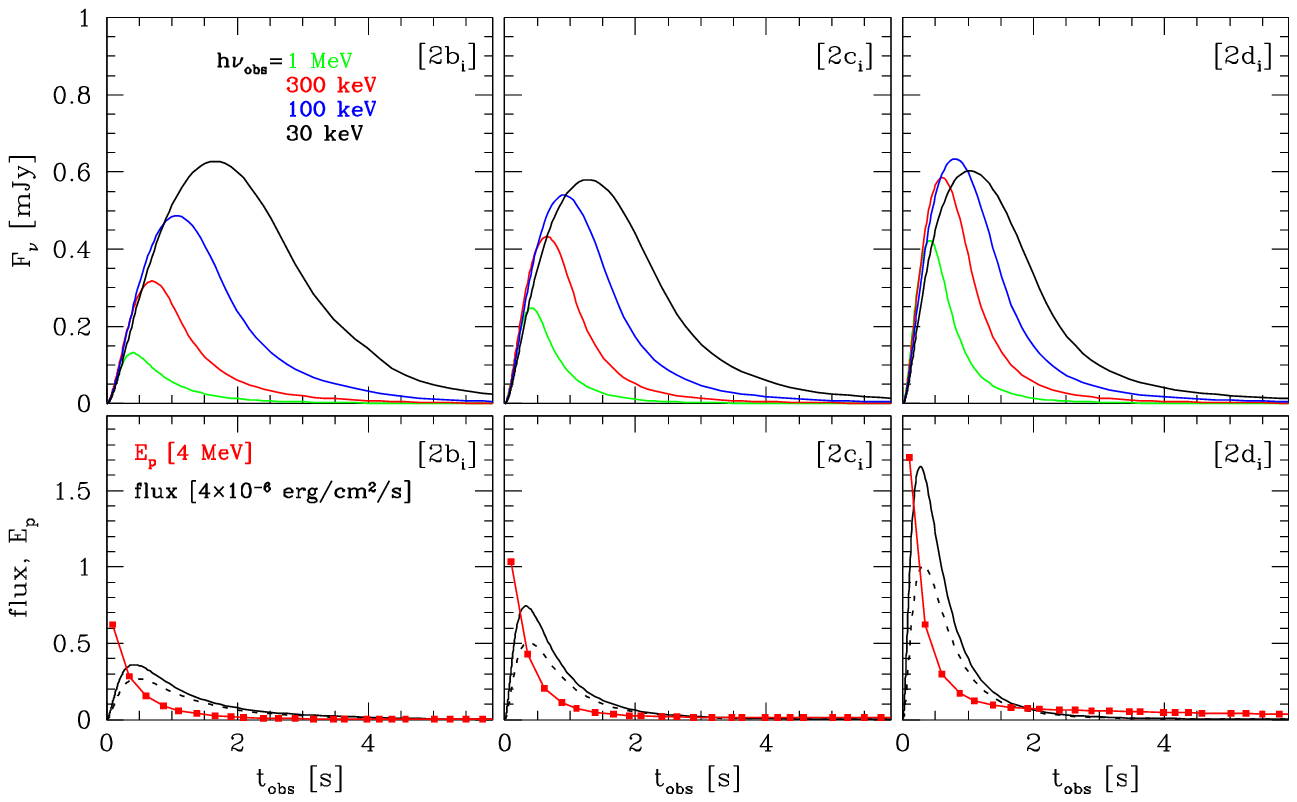}
\caption{
Same as in Figure~\ref{fig:f1}, but for the $i$ models [2b$_i$], [2c$_i$], and [2d$_i$].
}
\label{fig:f4}
\end{center}
\end{figure}

\begin{figure*} \centering
\centering
\begin{tabular}{cc}
\includegraphics[width=8.27cm]{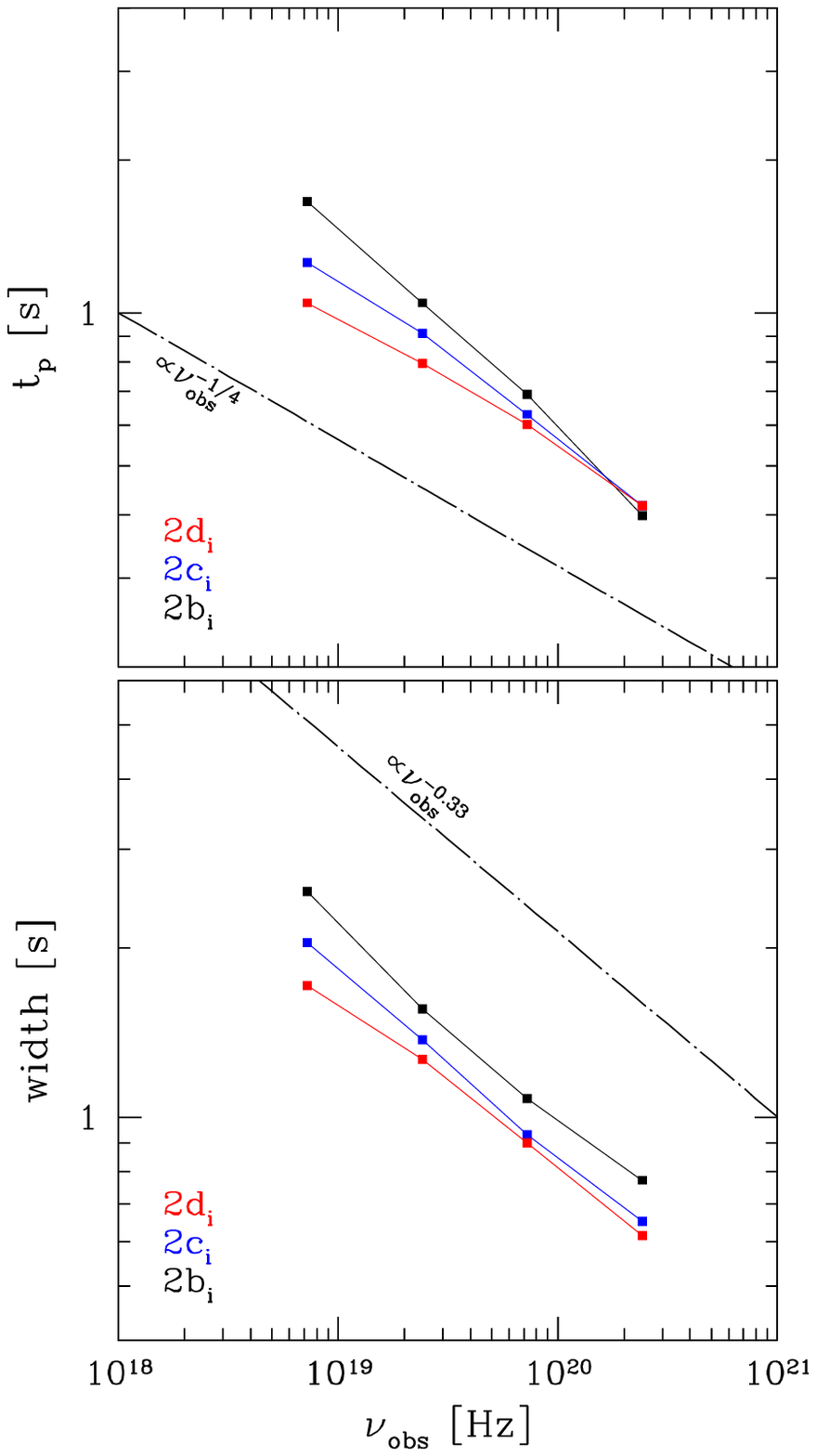} &
\includegraphics[width=8.50cm]{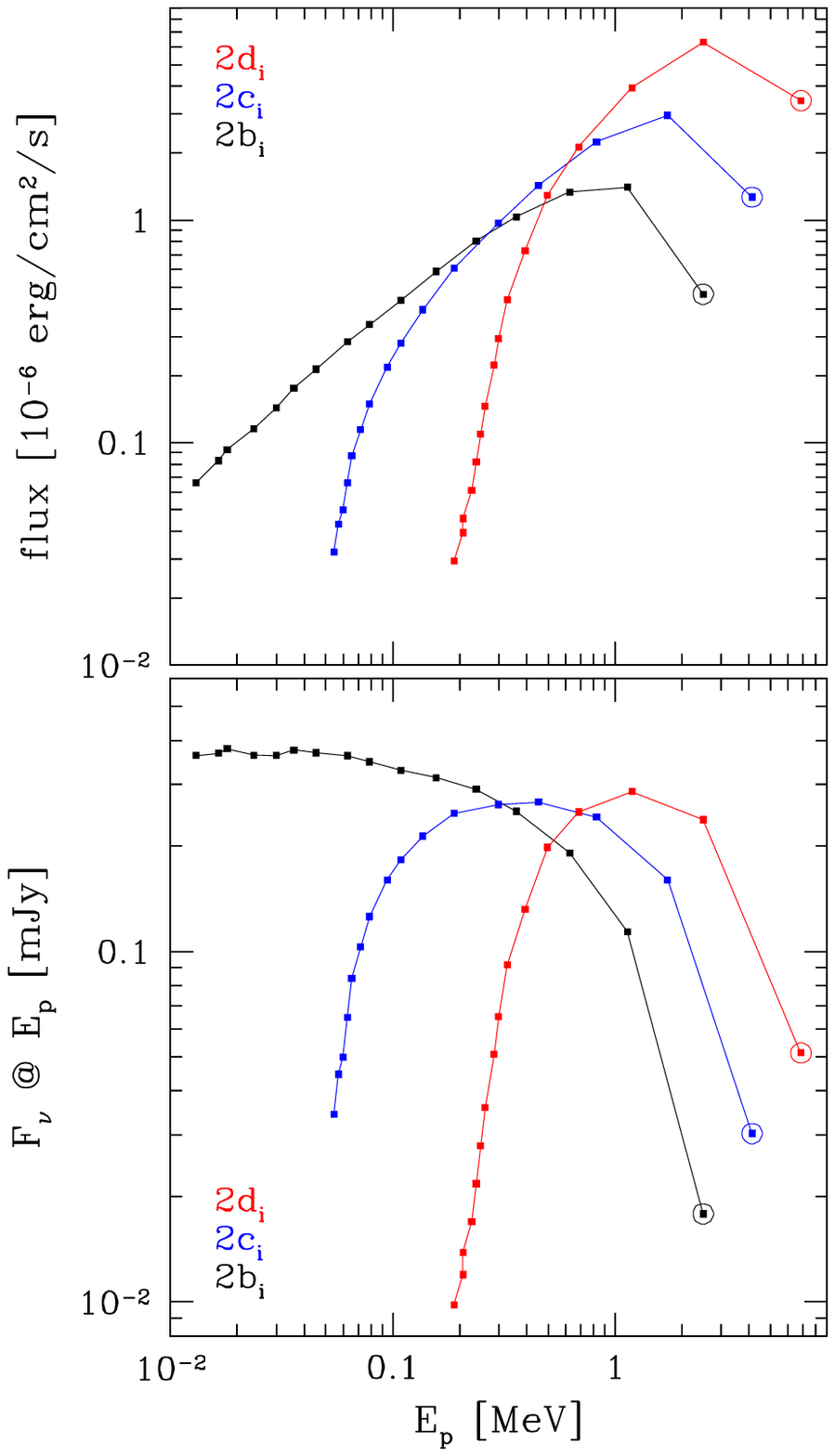} \\
\end{tabular}
\caption{
Same as in Figure~\ref{fig:f2}, but for the $i$ models [2b$_i$], [2c$_i$], and [2d$_i$].
}
\label{fig:f5}
\end{figure*}

Figure~\ref{fig:f4} shows the calculation results for the $i$ models [2b$_i$], [2c$_i$], and [2d$_i$], in which the four different light curves (top panels) and the temporal curves for the peak energy $E_p$ and the flux (bottom panels) are shown in the same way as in Figure~\ref{fig:f1}. In this case, the frequency $\nuobs$ along the observer's line of sight roughly follows $\nuobs \propto \Gamma B \gch^2 \propto r^{s-b+2g}$, and thus the peak energy $E_p$ decreases faster than in Figure~\ref{fig:f1}. As a result, the light curves form a broad pulse earlier than in Figure~\ref{fig:f1}, and the turning-off signature at about $\tobs=4$ s becomes nearly invisible. It is clear again that the $E_p$ temporal curve exhibits a hard-to-soft pattern while the light curves show the positive type of spectral lags. Figure~\ref{fig:f5} shows the properties of the $i$ models [2b$_i$], [2c$_i$], and [2d$_i$] in the same way as in Figure~\ref{fig:f2}. The peak time $t_p$ and the width of broad-pulse light curves, plotted in the left column, generally agree with the observations again. Also, a counter-clockwise pattern of the peak evolution is evident in each model, as shown in the right column. We remark here that a similar set of the results is to be obtained for other values of $g$ index as long as $g<0$.

\begin{figure}
\begin{center}
\includegraphics[width=18cm]{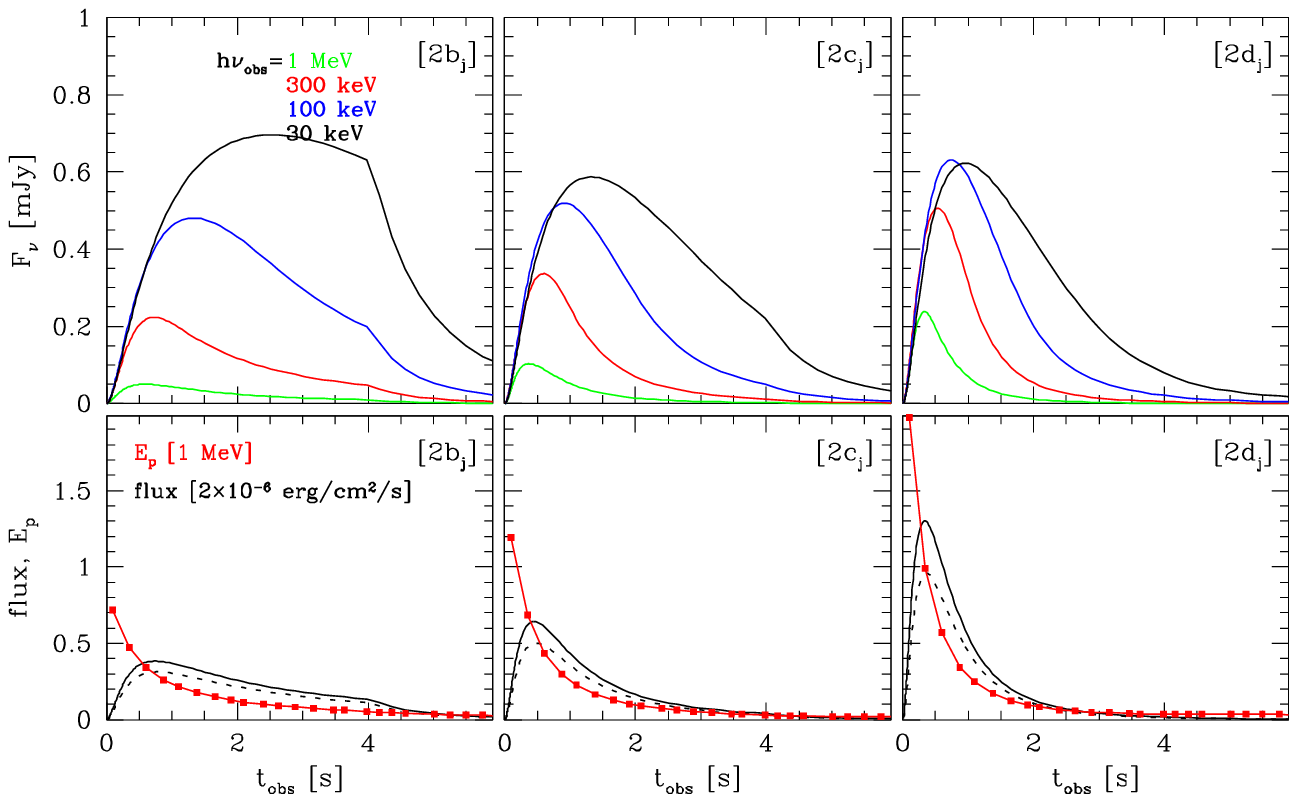}
\caption{
Same as in Figure~\ref{fig:f1}, but for the $j$ models [2b$_j$], [2c$_j$], and [2d$_j$].
}
\label{fig:f6}
\end{center}
\end{figure}

\begin{figure*} \centering
\centering
\begin{tabular}{cc}
\includegraphics[width=8.27cm]{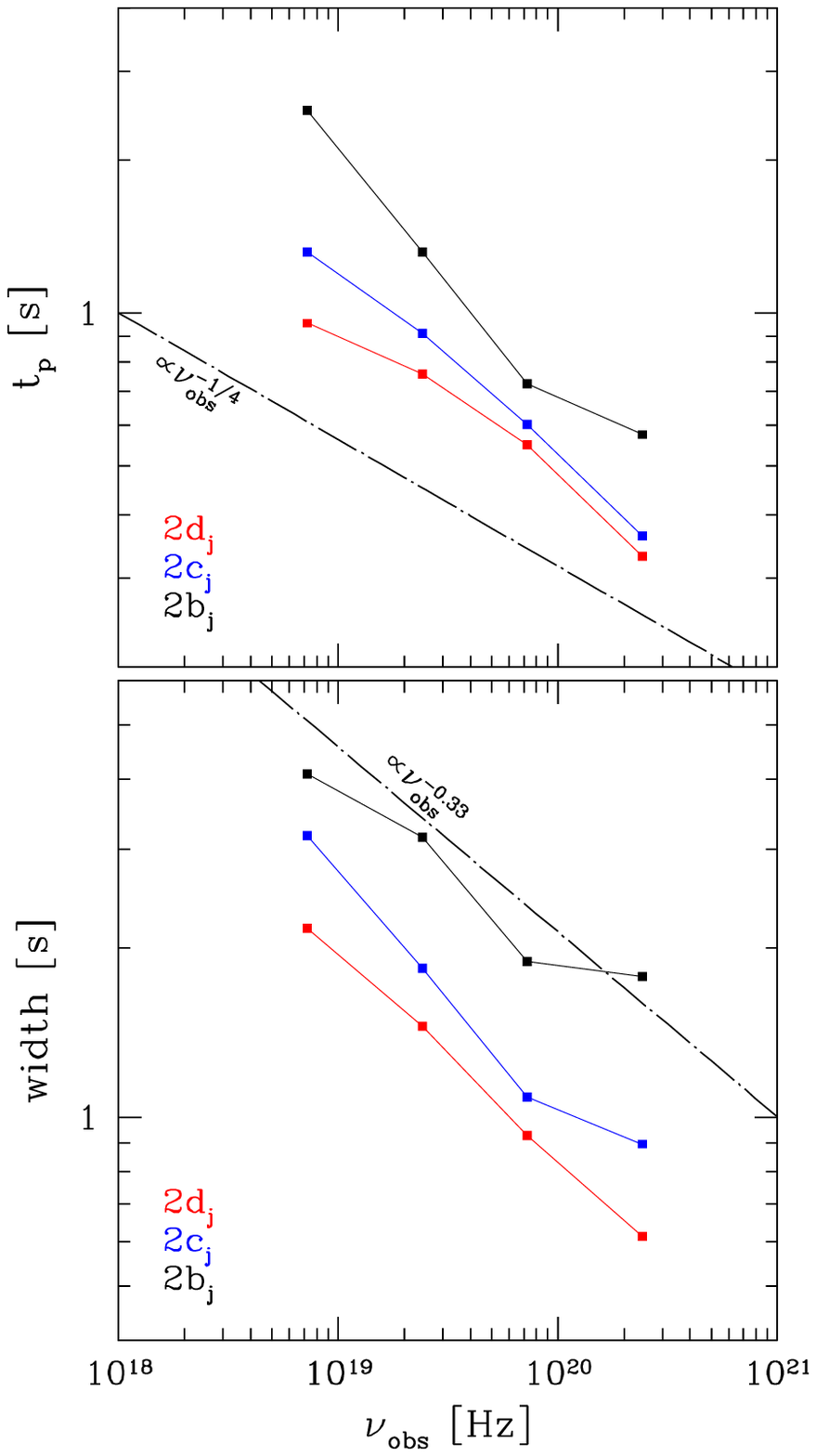} &
\includegraphics[width=8.50cm]{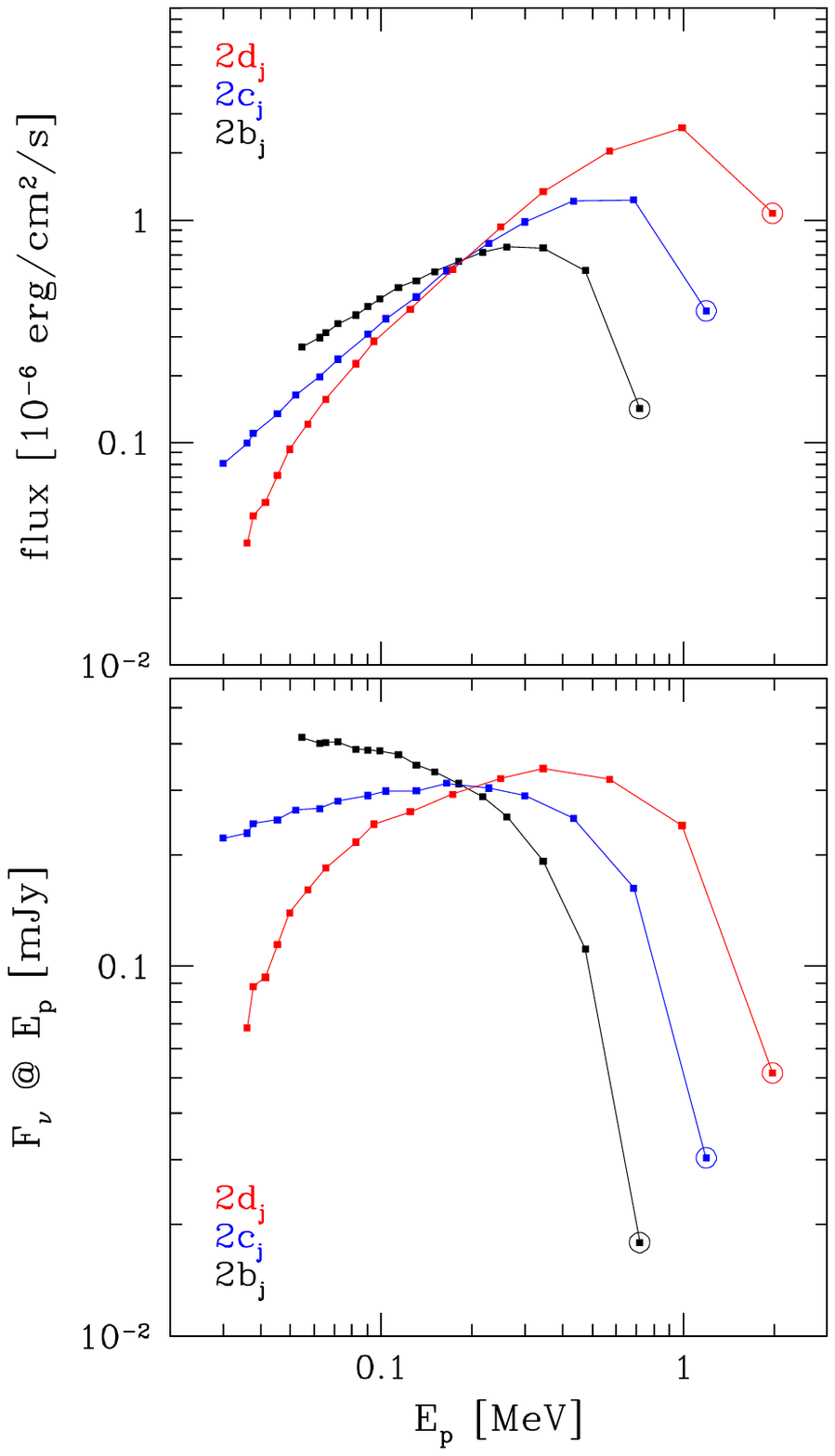} \\
\end{tabular}
\caption{
Same as in Figure~\ref{fig:f2}, but for the $j$ models [2b$_j$], [2c$_j$], and [2d$_j$].
}
\label{fig:f7}
\end{figure*}

Figures~\ref{fig:f6} and \ref{fig:f7} show the results for a new set of three models [2b$_j$], [2c$_j$], and [2d$_j$], whose model parameters are the same as in the models [2b$_i$], [2c$_i$], and [2d$_i$], respectively, except that a small, positive value of $g$ index with $g=0.1$ is taken. Overall, the obtained results are reasonably good and compatible with the observations, while exhibiting a clear hard-to-soft pattern of the $E_p$ evolution together with the positive type of spectral lags. However, since the peak energy $E_p$ here decreases slower than in Figure~\ref{fig:f1}, it becomes more difficult for the light curves to form a well-behaved broad pulse than in Figure~\ref{fig:f1}.

Now, in order to see a possibility of reproducing the flux-tracking pattern of $E_p$ evolution, we consider a broken power-law profile of $\gch(r)$ as follows
\beq
\label{eq:gch_broken}
\gch(r) = \gch^0 \times
\left\{
\begin{array}{ll}
(r/r_0)^g    \quad & \mbox{if} \quad r \leq r_0, \\
(r/r_0)^{-g} \quad & \mbox{if} \quad r \geq r_0,
\end{array}
\right.
\eeq 
where a normalization value $\gch^0$ is set to be $\gch^0 = 10^5$ at radius $r_0 = 10^{15}$ cm with an index $g = 0.5$. A broken power law function of $\gch$ may be possible when magnetic dissipation behavior changes at a critical radius $r_0$. For example, in the numerical simulations is \cite{deng15}, it is found that one ICMART event includes four different stages. Each stage involves different magnetic configurations, and may introduce slightly different behaviors of particle acceleration. Besides this $\gch$ profile, we keep all others the same as in the models [2b], [2c], and [2d] and have three new models [2b$_k$], [2c$_k$], and [2d$_k$], respectively.

\begin{figure}
\begin{center}
\includegraphics[width=18cm]{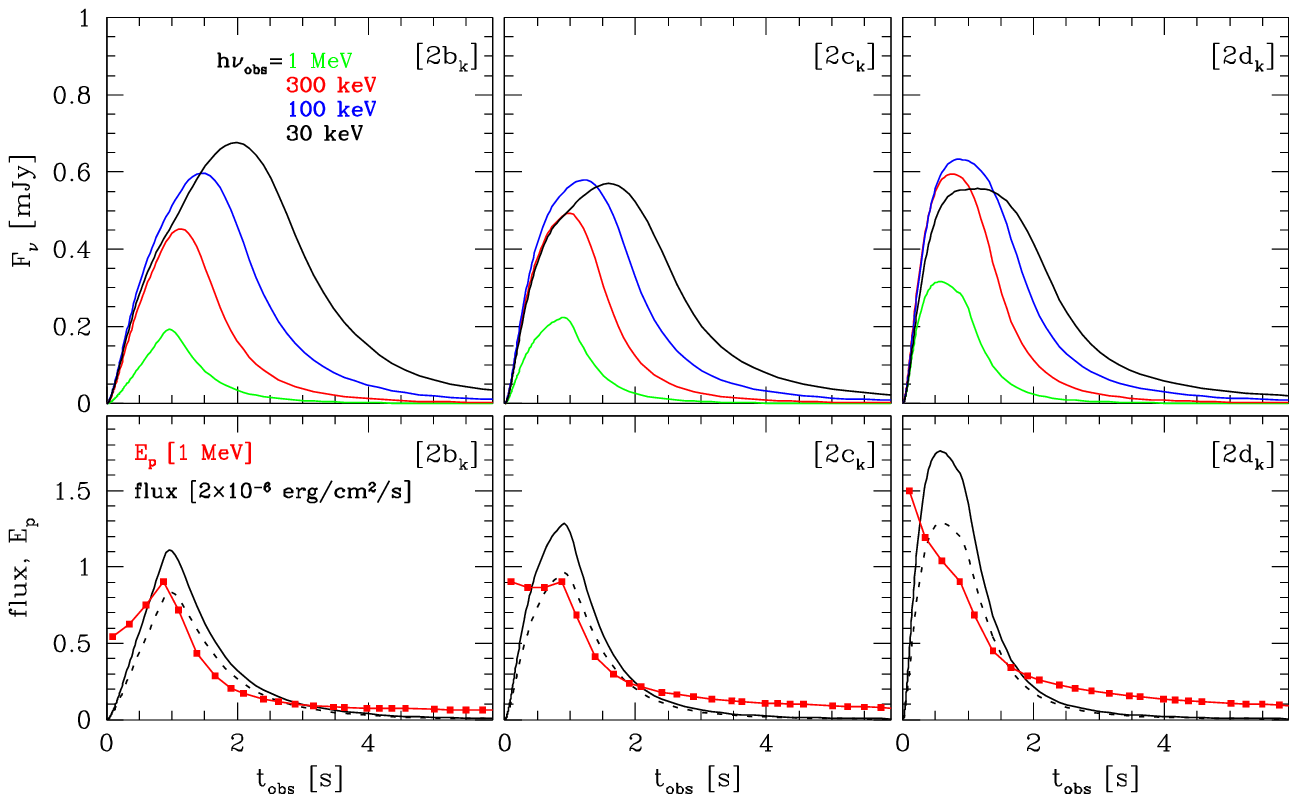}
\caption{
Same as in Figure~\ref{fig:f1}, but for the $k$ models [2b$_k$], [2c$_k$], and [2d$_k$].
}
\label{fig:f8}
\end{center}
\end{figure}

\begin{figure*} \centering
\centering
\begin{tabular}{cc}
\includegraphics[width=8.27cm]{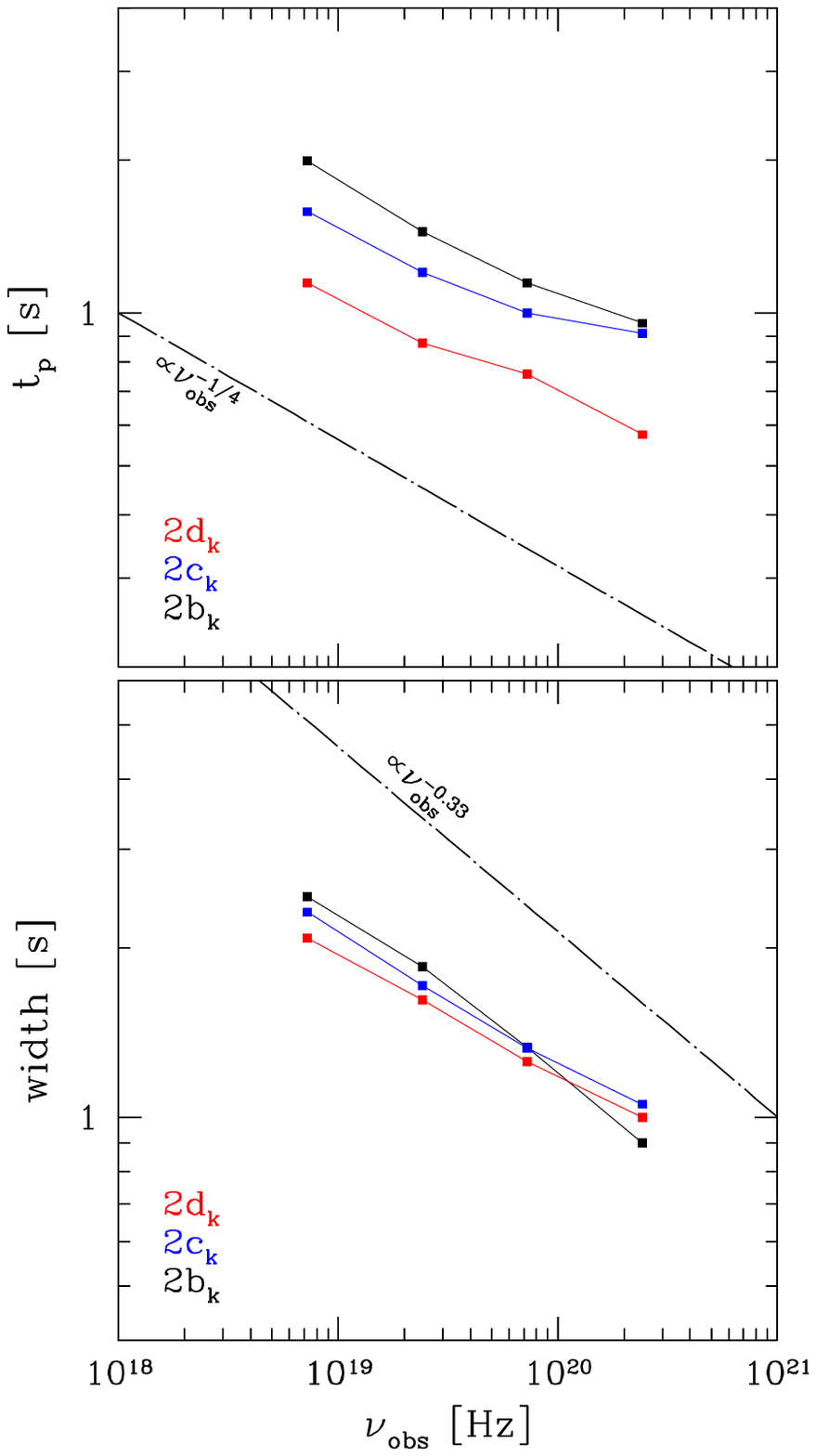} &
\includegraphics[width=8.50cm]{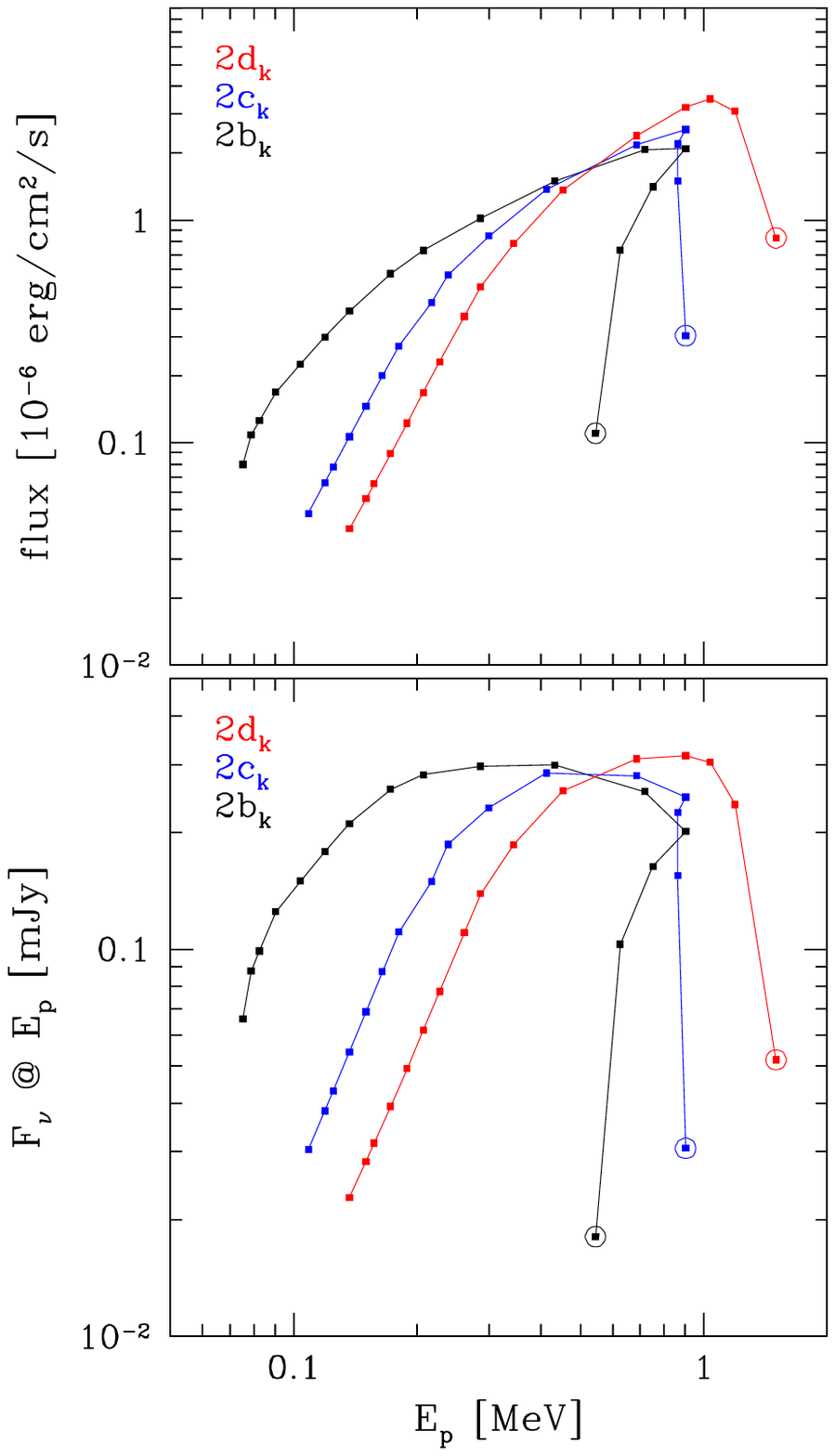} \\
\end{tabular}
\caption{
Same as in Figure~\ref{fig:f2}, but for the $k$ models [2b$_k$], [2c$_k$], and [2d$_k$].
}
\label{fig:f9}
\end{figure*}

Figures~\ref{fig:f8} and \ref{fig:f9} show the results of the $k$ models [2b$_k$], [2c$_k$], and [2d$_k$]. Firstly, we note that we indeed have a flux-tracking pattern of $E_p$ evolution here in the model [2b$_k$]. Also, interestingly, we have three different types of the peak evolution in this case; namely, we have a hardening, flattening, and softening pattern, seen in the models [2b$_k$], [2c$_k$], and [2d$_k$], respectively, during the rising phase of the flux curve. This can be understood by recalling that the frequency $\nuobs$ along the observer's line of sight roughly follows $\nuobs \propto r^{s-b+2g}$ when $r \leq r_0$, and $s-b+2g$ = (0.35, 0.1, -0.15) for ([2b$_k$], [2c$_k$], [2d$_k$]), respectively. In all three models, the light curves exhibit the positive type of spectral lags, with the broad-pulse properties well compatible with the observations. The lower right panel in Figure~\ref{fig:f9} clearly displays a counter-clockwise pattern of the peak evolution in each model.

\begin{figure}
\begin{center}
\includegraphics[width=18cm]{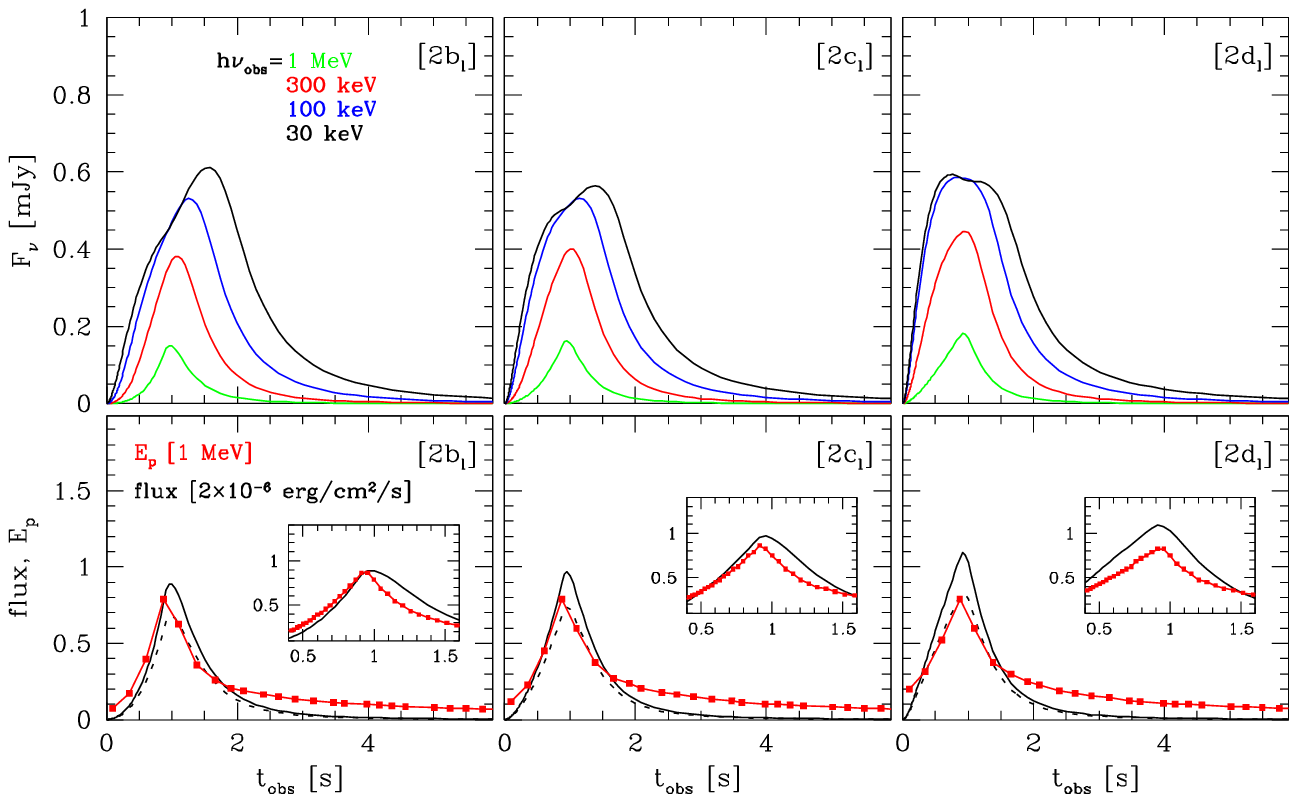}
\caption{
Same as in Figure~\ref{fig:f1}, but for the $l$ models [2b$_l$], [2c$_l$], and [2d$_l$]. The insets in the bottom panels show a zoom-in view around the peak area of $E_p$ and flux curves.
}
\label{fig:f10}
\end{center}
\end{figure}

\begin{figure*} \centering
\centering
\begin{tabular}{cc}
\includegraphics[width=8.27cm]{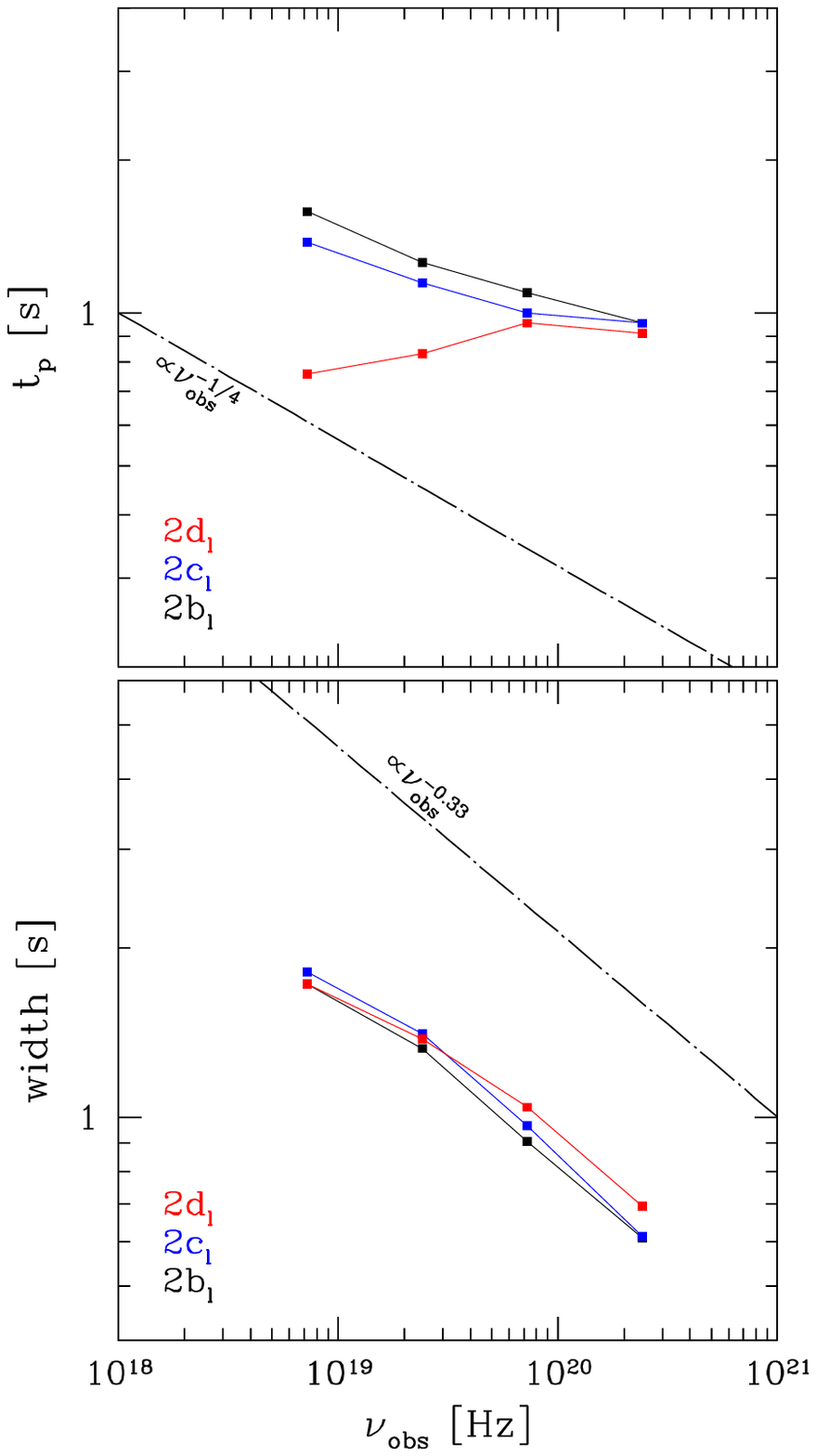} &
\includegraphics[width=8.50cm]{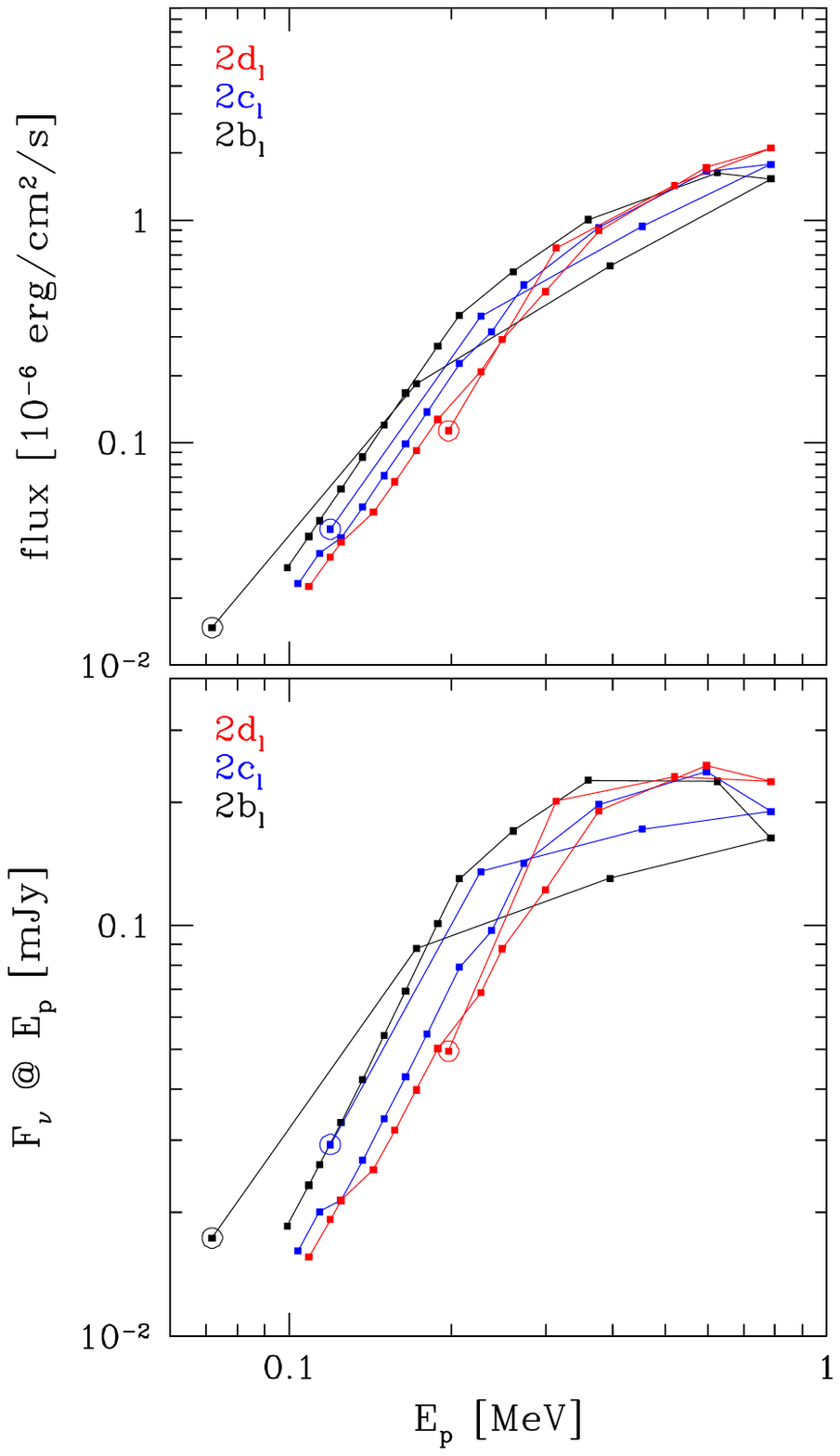} \\
\end{tabular}
\caption{
Same as in Figure~\ref{fig:f2}, but for the $l$ models [2b$_l$], [2c$_l$], and [2d$_l$].
}
\label{fig:f11}
\end{figure*}

It is then clear that we can reproduce more examples showing the flux-tracking pattern by increasing the value of $g$ index in Equation (\ref{eq:gch_broken}). We replace the $g$ index in Equation (\ref{eq:gch_broken}) by $g=1.0$ and form three new models named as [2b$_l$], [2c$_l$], and [2d$_l$], respectively. The results of the $l$ models are shown in Figures~\ref{fig:f10} and \ref{fig:f11}. As one can see, a flux-tracking pattern of $E_p$ evolution is firmly reproduced in all three models. The light curves in the models [2b$_l$] and [2c$_l$] still show the positive type of spectral lags. However, the light curves in the model [2d$_l$] exhibit a hint on the opposite pattern, i.e., the negative type of spectral lags. This can also be noticed, in the upper left panel of Figure~\ref{fig:f11}, by a positive slope for the model [2d$_l$]. The width properties of the broad pulses are in a good agreement with the observations in all three models. We now point out that there exists an important difference between the positive- and the negative-type of spectral lags. Since the ($E_p$, flux) points in the upper right panel of Figure~\ref{fig:f11} are populated too densely, we ask the readers to look at the ($E_p$, $F_{\nu, E_p}$) points in the lower right panel of Figure~\ref{fig:f11}. For the models [2b$_l$] and [2c$_l$] with the positive type of spectral lags, we still have a counter-clockwise pattern of the peak evolution (with a self-crossing in its pattern curve this time). However, for the model [2d$_l$] with the negative type of spectral lags, we have an evolving curve that starts to show a clockwise pattern rather than a counter-clockwise pattern; we will present better examples below regarding this point. Furthermore, we find another difference between the positive- and the negative-type of spectral lags by closely looking at the peak area of $E_p$ and flux curves. The insets inserted in the bottom panels of Figure~\ref{fig:f10} show a zoom-in plot around the peak area. As one can see, for the models with the positive type of spectral lags, the peak location of the flux curve slightly lags behind the peak of $E_p$ curve. On the other hand, for the model with the negative type of spectral lags, there is no longer a visible lag between the two curves.

\begin{figure}
\begin{center}
\includegraphics[width=18cm]{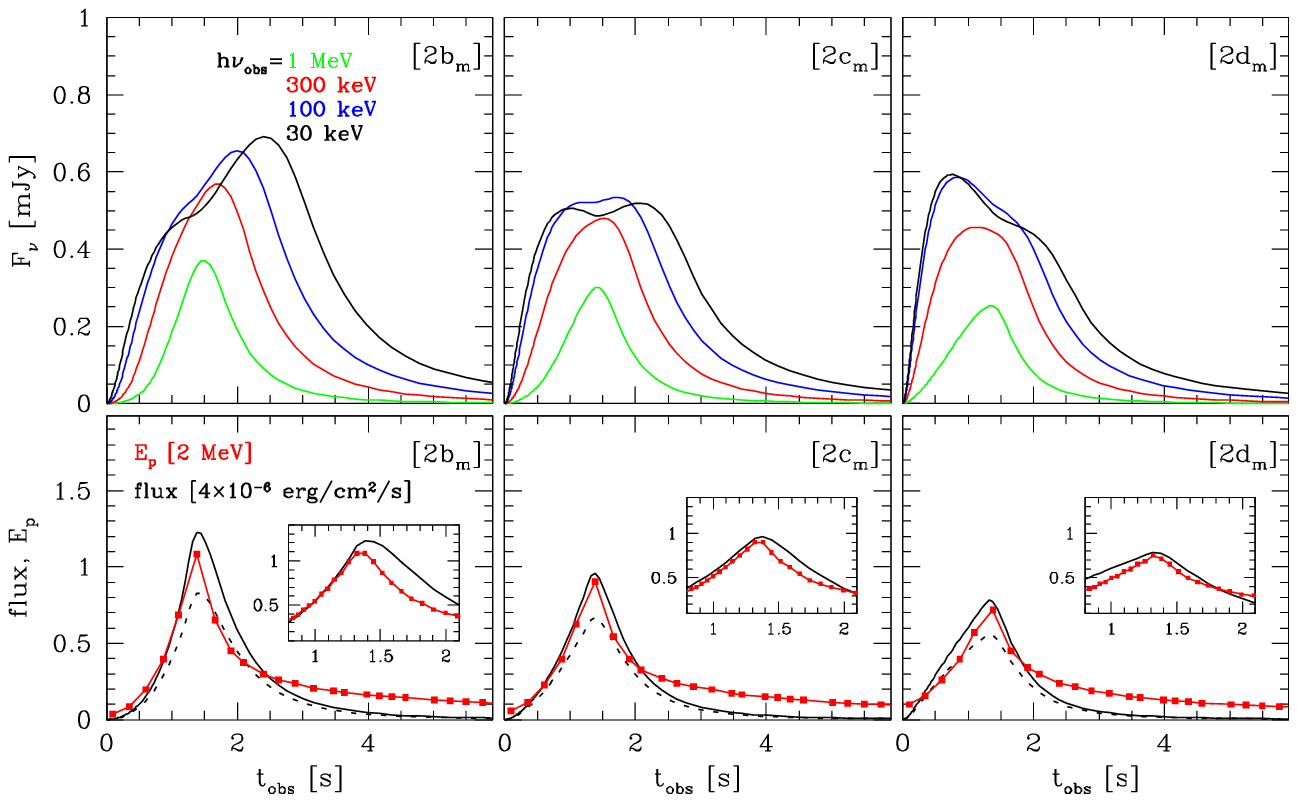}
\caption{
Same as in Figure~\ref{fig:f1}, but for the $m$ models [2b$_m$], [2c$_m$], and [2d$_m$]. The insets in the bottom panels show a zoom-in view around the peak area of $E_p$ and flux curves.
}
\label{fig:f12}
\end{center}
\end{figure}

\begin{figure*} \centering
\centering
\begin{tabular}{cc}
\includegraphics[width=8.27cm]{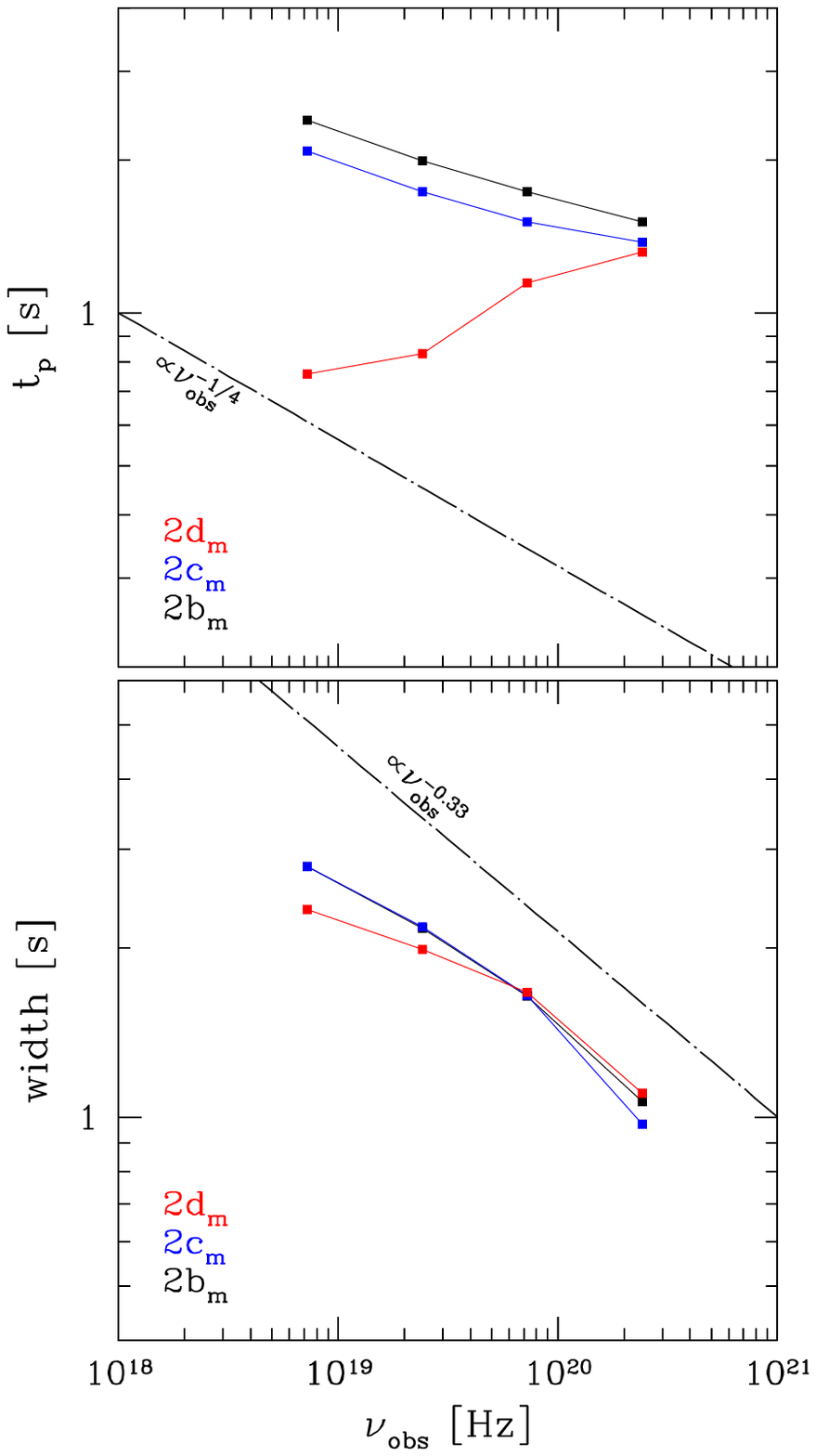} &
\includegraphics[width=8.50cm]{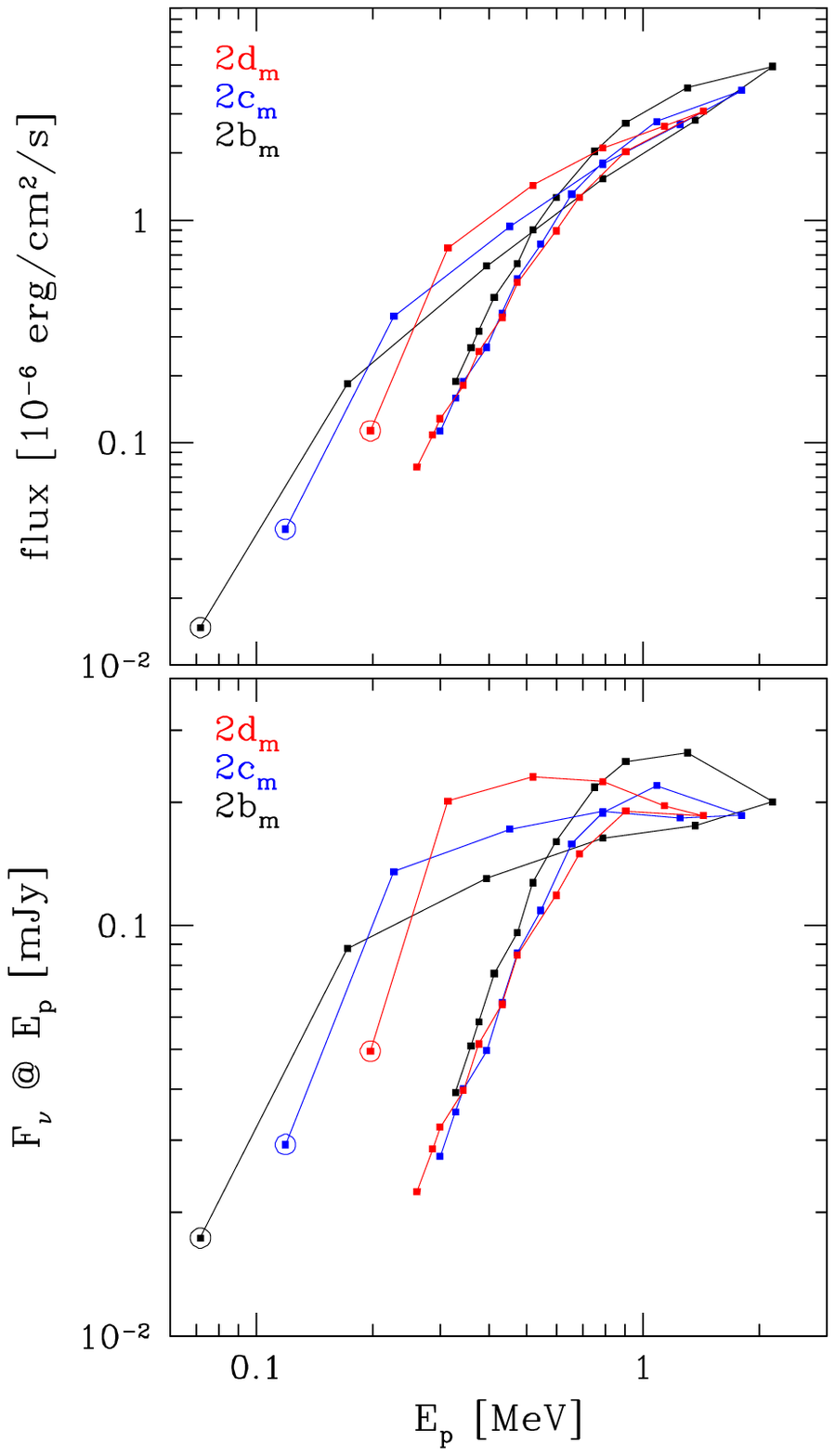} \\
\end{tabular}
\caption{
Same as in Figure~\ref{fig:f2}, but for the $m$ models [2b$_m$], [2c$_m$], and [2d$_m$].
}
\label{fig:f13}
\end{figure*}

Figures~\ref{fig:f12} and \ref{fig:f13} show the results of three new models, called [2b$_m$], [2c$_m$], and [2d$_m$], whose model parameters are the same as in the models [2b$_l$], [2c$_l$], and [2d$_l$], respectively, except that Equation (\ref{eq:gch_broken}) has $\gch^0 = 2 \times 10^5$ and $r_0 = 2 \times 10^{15}$ cm. The $\gch$ profile for the $m$ models is identical to that of the $l$ models when $r \leq 10^{15}$ cm, and then it extends further up to a higher value than in the $l$ models (see Figure~\ref{fig:f3}). In all three $m$ models, we have a strong flux-tracking pattern of $E_p$ evolution. While the light curves in the models [2b$_m$] and [2c$_m$] show the positive type of spectral lags, the light curves in the model [2d$_m$] exhibit, very clearly this time, the negative type of spectral lags. This can also be seen in the upper left panel of Figure~\ref{fig:f13}. Once again, in the right column of Figure~\ref{fig:f13}, it is more useful for the readers to look at the ($E_p$, $F_{\nu, E_p}$) points than the ($E_p$, flux) points, in order to understand and differentiate the characteristics of the models. It is clear that the model [2d$_m$] with the negative type of spectral lags shows a clockwise pattern in its peak-evolving curve whereas the models [2b$_m$] and [2c$_m$] with the positive type of spectral lags show a counter-clockwise pattern of the peak evolution with a self-crossing in their pattern curve. Also, as shown in the insets inserted in the bottom panels of Figure~\ref{fig:f12}, the models with the positive type of spectral lags have a flux curve that slightly lags behind the $E_p$ curve in their peaking time. On the other hand, the model with the negative type of spectral lags does not show a visible lag between the two curves.

\begin{figure}
\begin{center}
\includegraphics[width=18cm]{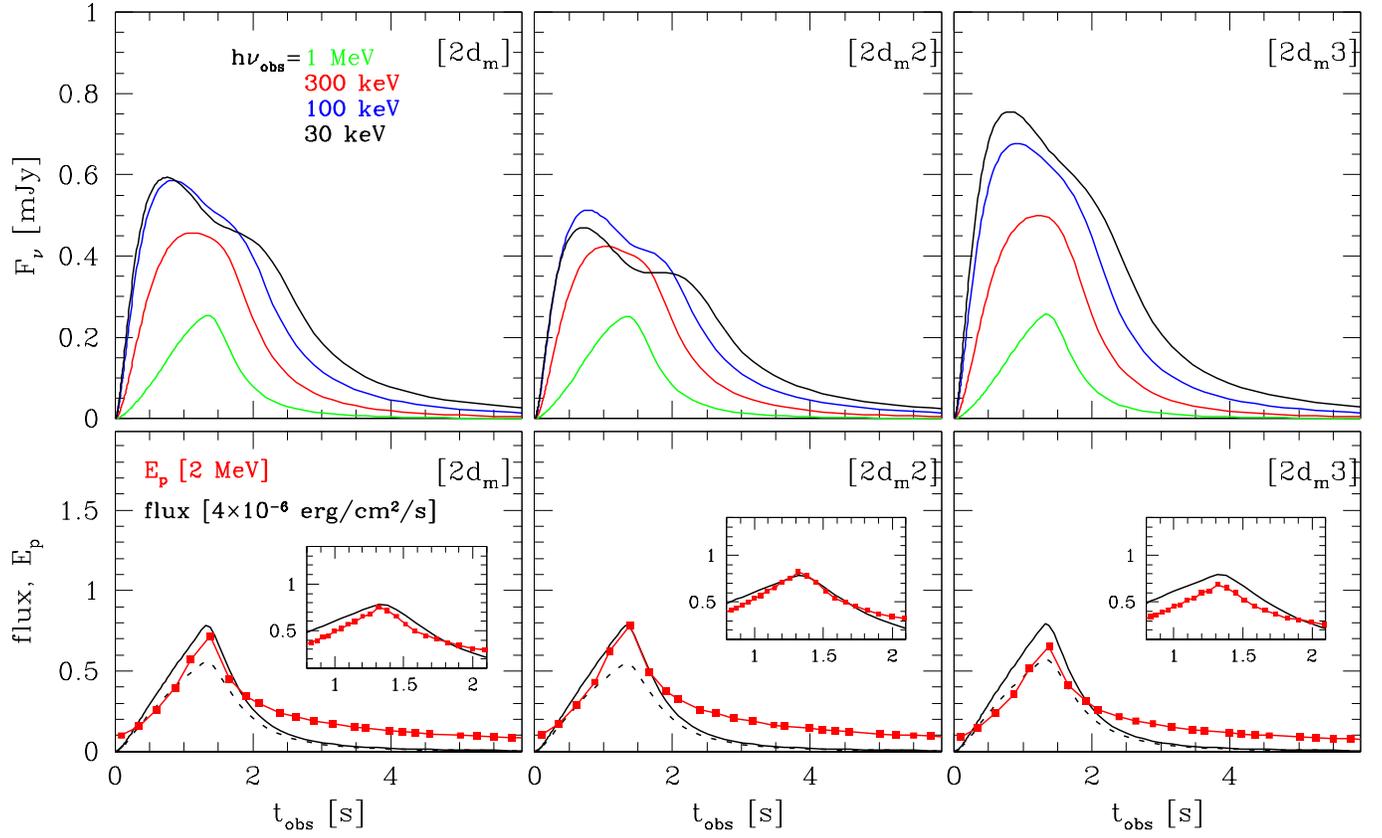}
\caption{
Same as in Figure~\ref{fig:f1}, but for the $m$ models [2d$_m$], [2d$_m$2], and [2d$_m$3]. These three models are identical to one another except for the Band-function $\alphaB$ index; see the text. The insets in the bottom panels show a zoom-in view around the peak area of $E_p$ and flux curves.
}
\label{fig:f14}
\end{center}
\end{figure}

\begin{figure*} \centering
\centering
\begin{tabular}{cc}
\includegraphics[width=8.27cm]{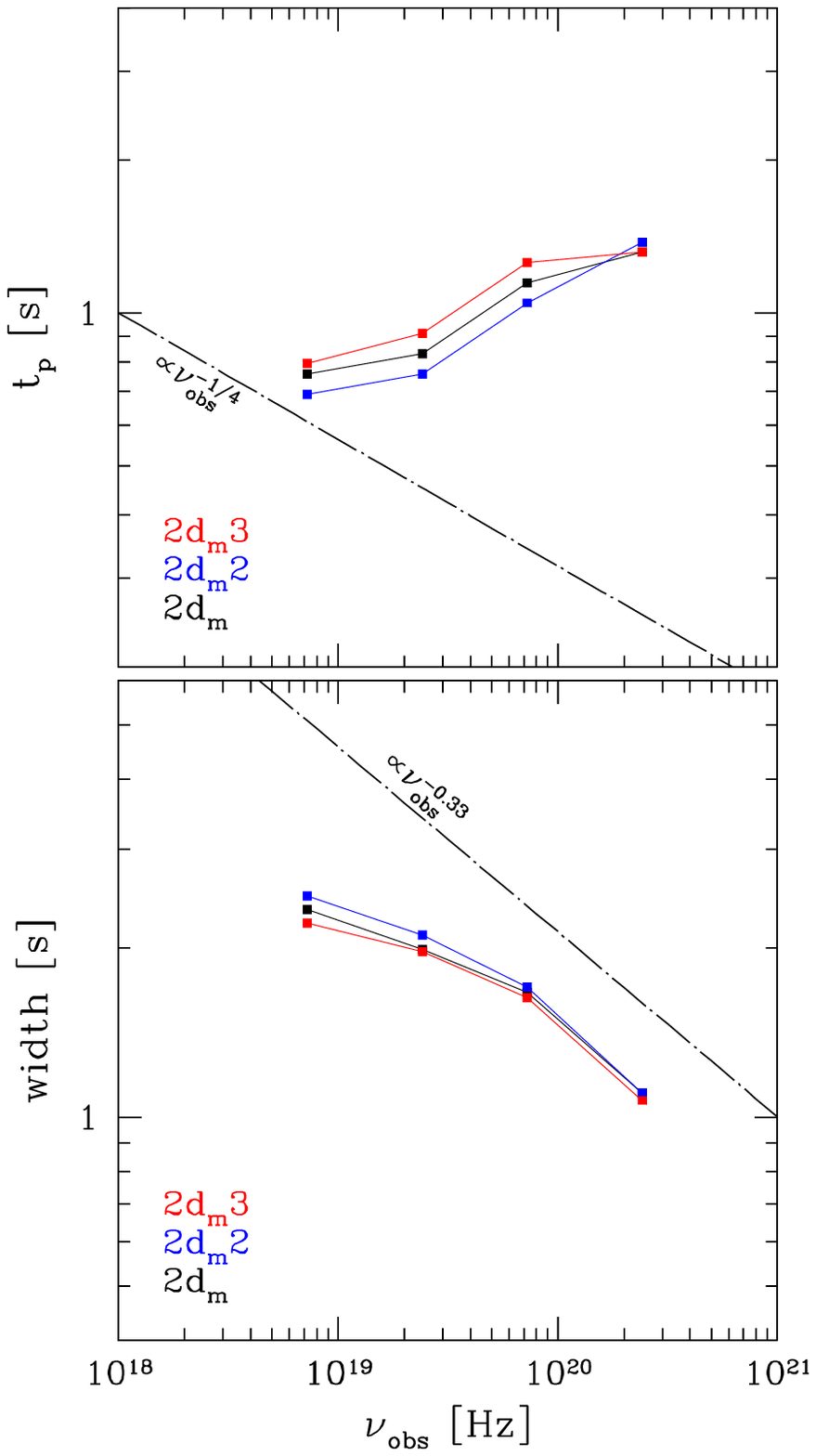} &
\includegraphics[width=8.50cm]{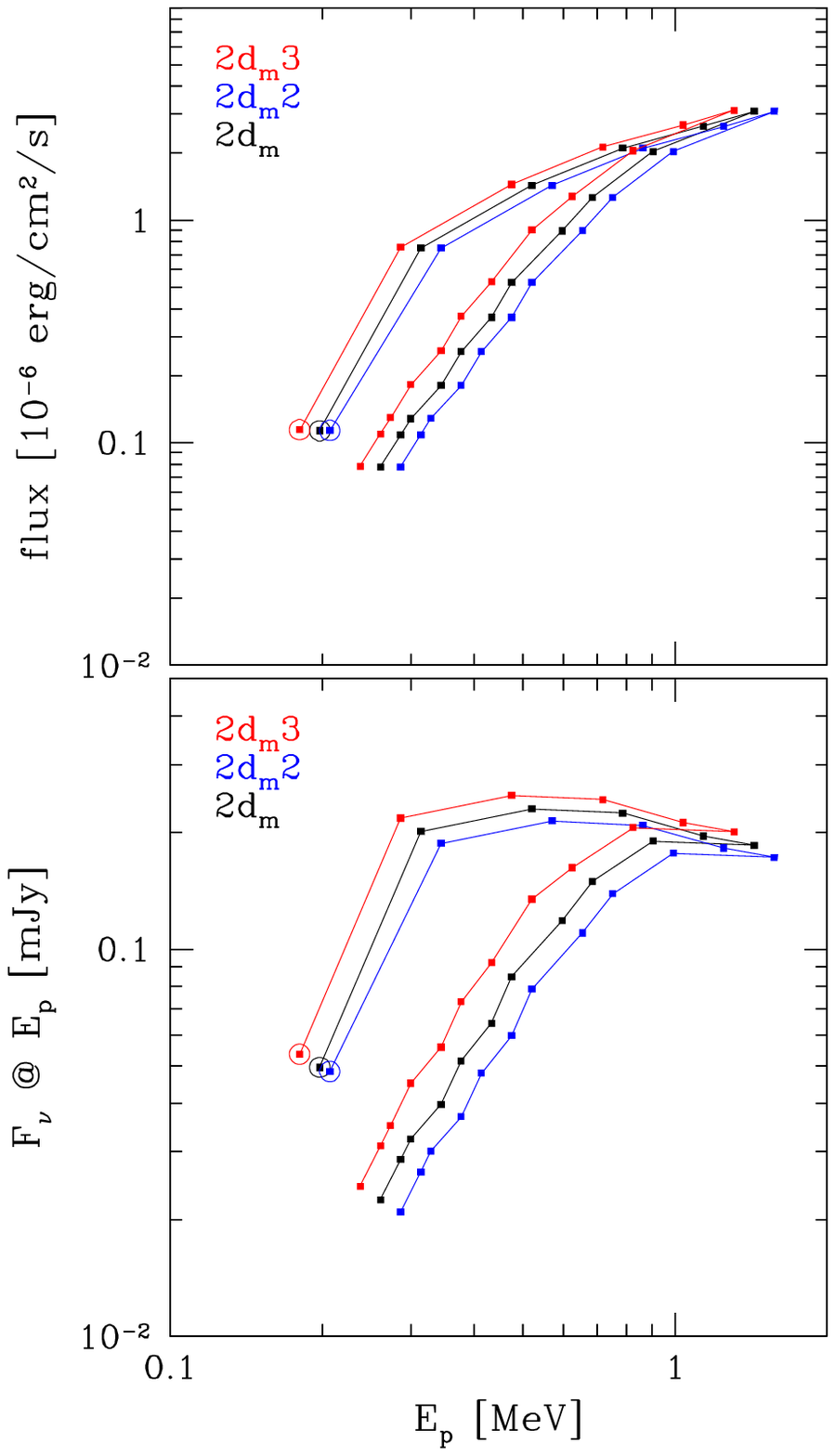} \\
\end{tabular}
\caption{
Same as in Figure~\ref{fig:f2}, but for the $m$ models [2d$_m$], [2d$_m$2], and [2d$_m$3]. These three models are identical to one another except for the Band-function $\alphaB$ index.
}
\label{fig:f15}
\end{figure*}

Another interesting thing that we note from the light curves of the $m$ models in Figure~\ref{fig:f12} is that the ``double-peaked'' broad pulses are chromatically present in the low-energy curves. We now demonstrate that this double-peaked feature depends sensitively on the Band-function $\alphaB$ index that we use to describe the functional form $H(x)$ in the co-moving frame. We take the model [2d$_m$], as an example, whose $\alphaB$ index is $-0.8$, and form two new models [2d$_m$2] and [2d$_m$3] by replacing the $\alphaB$ index by $-0.7$ and $-0.9$, respectively. The result is shown in Figure~\ref{fig:f14}. As one can see, the harder the $\alphaB$ index is, the stronger the double-peaked feature is. A flux-tracking pattern of $E_p$ evolution and the negative type of spectral lags still remain in the new models [2d$_m$2] and [2d$_m$3]. It is clear in Figure~\ref{fig:f15} that these models have the negative type of spectral lags with a clockwise pattern curve of the peak evolution.

%--------------------------------------------------------
%
% Section 4
%
%--------------------------------------------------------

\section{Conclusions and Discussion} \label{section:4}

In this paper, we consider a simple physical picture, in which a thin relativistic spherical shell expands in space radially while emitting radiation uniformly from all locations in the shell. An isotropic angular distribution of the emitted power is also assumed in the co-moving frame of the shell. We take fully into account the high-latitude emission effect of the spherical shell, by making use of the formulation given in \cite{uhm15}, and calculate the observed spectral flux as a function of the observer time $\tobs$ and the observed frequency $\nuobs$. Following the findings shown in the first paper \citep{uhm16b}, we firstly take a Band-function shape to describe the emission spectrum H(x) in the co-moving frame and invoke synchrotron radiation to give the characteristic frequency and the spectral power of the electrons. Then we consider a globally-decreasing strength of magnetic fields, $B(r) \propto r^{-b}$, in the co-moving frame of the shell, with three different $b$ indices 1.0, 1.25, and 1.5 for the b-, c-, and d-models, respectively. Also, we let the emitting region itself undergo bulk acceleration by using an increasing profile of the Lorentz factor, $\Gamma(r) \propto r^s$, with the index $s=0.35$.

\begin{figure}
\begin{center}
\includegraphics[width=10cm]{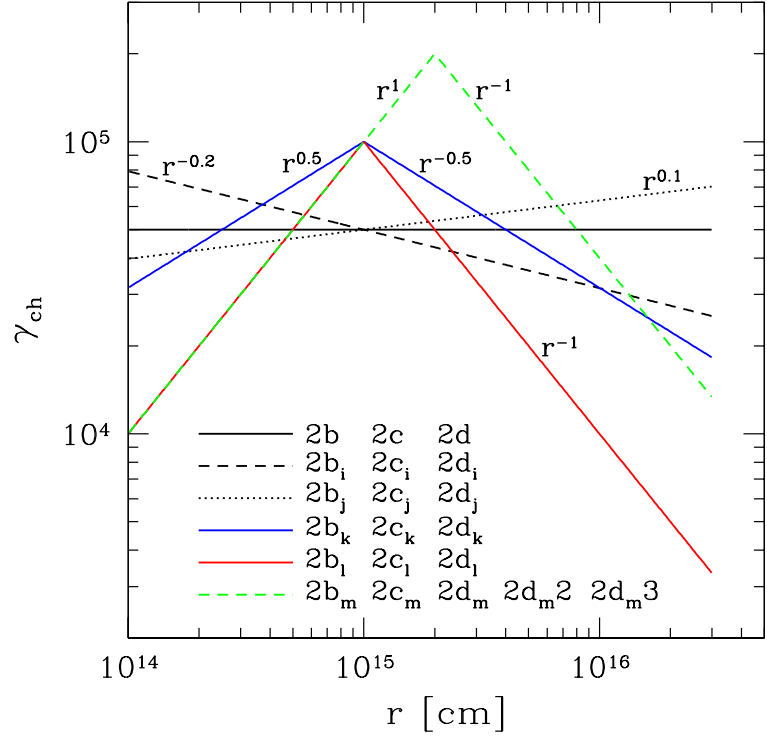}
\caption{
Profiles of characteristic Lorentz factor $\gch$ of electrons in the shell, for the twenty numerical models presented in the paper. The first three models [2b], [2c], and [2d] have a constant value of $\gch=5 \times 10^4$. Then we explore five different variations on the $\gch$ profile, which are indicated by five subscripts $i$, $j$, $k$, $l$, and $m$ contained in the model names. As in the models [2b], [2c], and [2d], the letters `b', `c', and `d' included in a model name always indicate for the $b$ index 1.0, 1.25, and 1.5, respectively, for a globally-decreasing strength of magnetic fields, $B(r) \propto r^{-b}$. The number `2' in the beginning of all model names is to indicate that the emitting region of these models undergoes bulk acceleration, $\Gamma(r) \propto r^s$, and one single index $s=0.35$ is used in all twenty models shown here. Lastly, unless an additional number is added at the end of a model name, the low-energy photon index $\alphaB$ of the Band-function shape is taken to be $-0.8$. The two models [2d$_m$2] and [2d$_m$3] have the $\alphaB$ index $-0.7$ and $-0.9$, respectively.
}
\label{fig:f3}
\end{center}
\end{figure}

Since there is no concrete prediction of the electron characteristic Lorentz factor $\gch(r)$ in the shell from the first principles, we explore a variety of analytical $\gch(r)$ profiles, as shown in Figure~\ref{fig:f3}, and show that the two distinct patterns of the peak-energy ($E_p$) evolution, i.e., the hard-to-soft and the flux-tracking behavior, are successfully and clearly reproduced in the results of our numerical models, just as revealed by the observations of broad pulses in the prompt phase of GRBs. Also, we show that the two different (i.e., the positive and the negative) types of spectral lags are successfully reproduced in the broad-pulse light curves of our numerical models. We stress that this is the first time that all these intriguing observational features, seen in the prompt gamma-rays of GRBs, are successfully reproduced within a physically-motivated model\footnote{There have been previous efforts of fitting the data using the physically-motivated models (e.g. using the curvature effect, \citealt{ryde02,kocevski03}), but these models did not consider the details of particle acceleration and synchrotron radiation.}. We further show that the patterns of the $E_p$ evolution have, in fact, close connections to the occurrence of the positive and the negative type of spectral lags. In particular, we find the followings:
\begin{itemize}
\item Only the positive type of spectral lags can occur in the case of a hard-to-soft evolution of the peak energy;
\item Both the positive and the negative types of spectral lags can occur in the case of a flux-tracking pattern of the peak energy;
\item A time-evolving curve of showing the ($E_p$, $F_{\nu, E_p}$) points, which describes the peak evolution, exhibits a counter-clockwise pattern for the positive type of spectral lags, but a clockwise pattern for the negative type of spectral lags;
\item For the flux-tracking pattern, the peak location of the flux curve slightly lags behind the peak of $E_p$ curve if the spectral lags are positive, whereas there is no longer a visible lag between the two curves if the spectral lags are negative;
\item For the flux-tracking pattern, double-peaked broad pulses can chromatically appear in the low-energy light curves. The harder the low-energy photon index $\alphaB$ of the Band-function shape, the stronger the double-peaked feature.
\end{itemize}

These points may be understood intuitively. Here we have a curved shape for the emission spectrum $H(x)$. In the case of a counter-clockwise pattern, the observed spectrum sweeps through the observer energy-space in a counter-clockwise manner, as represented by the spectral flux $F_{\nu, E_p}$ at the peak energy $E_p$, and therefore the observed spectral flux gets larger at higher energy first and then progressively at lower energy later, hence resulting in the positive type of spectral lags. On the other hand, in the case of a clockwise pattern, it happens in the opposite way, thus leading to the negative type of spectral lags.

For the hard-to-soft behavior of the peak energy, it is only possible to have a counter-clockwise pattern since the peak energy should always decrease while the flux rises and then falls. Therefore, only the positive type of spectral lags is expected to be possible. For the flux-tracking behavior of the peak energy, both a counter-clockwise and a clockwise pattern of the peak evolution is plausible depending on the physical parameters, and thus we have both the positive and the negative type of spectral lags.

The numerical models presented in this paper have three different values for the index $b$ and invoke for many different $\gch$ profiles, in order to explore diverse patterns of the peak evolution and to reproduce all those intriguing observational features, mentioned above. Nevertheless, it appears that the properties of broad-pulse light curves, in particular, the width relations of broad pulses, remain compatible with the observations for all the numerical models presented here. This strongly suggests that the $s$ index (of showing the bulk acceleration) is probably the ``main shaper'' of the pulse properties. Also, as we stressed in the first paper \citep{uhm16b}, this requirement of bulk acceleration provides a ``smoking-gun'' evidence for a significant Poynting flux carried by relativistic jets in GRBs.

%--------------------------------------------------------
%
% Acknowledgments
%
%--------------------------------------------------------

\acknowledgments
This research was supported by an appointment to the NASA Postdoctoral Program at the Goddard Space Flight Center, administered by Universities Space Research Association through a contract with NASA.
This work was also supported by NASA through an Astrophysical Theory Program (grant number NNX 15AK85G) and an Astrophysics Data Analysis Program (grant number NNX 14AF85G).

%--------------------------------------------------------
%
% References
%
%--------------------------------------------------------

%\bibliography{ms}

%--------------------------------------------------------
%
% Tables
%
%--------------------------------------------------------

\clearpage

%--------------------------------------------------------
\begin{deluxetable}{lllccccr}
\tabletypesize{\tiny}
\tablewidth{0pt}
\tablecaption{Model parameters of our numerical models}
%\tablecolumns{7}
\tablehead{
\colhead{Model Name$\,$\tablenotemark{a}}&
\colhead{Index $s$$\,$\tablenotemark{b}}&
\colhead{Index $b$$\,$\tablenotemark{c}}&
\colhead{$\gch$ Profile$\,$\tablenotemark{d}}&
\colhead{Index $g$$\,$\tablenotemark{e}}&
\colhead{$\gch^0$$\,$\tablenotemark{f}}&
\colhead{$r_0$ for $\gch$ (cm)$\,$\tablenotemark{g}}&
\colhead{$\alphaB$$\,$\tablenotemark{h}} 
}
\startdata
2b       &  0.35  & 1.0   & const                    &  --   &  $5\times 10^4$  &  --                 &  -0.8  \\
2c       &  0.35  & 1.25  & const                    &  --   &  $5\times 10^4$  &  --                 &  -0.8  \\
2d       &  0.35  & 1.5   & const                    &  --   &  $5\times 10^4$  &  --                 &  -0.8  \\
2b$_i$   &  0.35  & 1.0   & Eq (\ref{eq:gch})        & -0.2  &  $5\times 10^4$  &  $10^{15}$          &  -0.8  \\
2c$_i$   &  0.35  & 1.25  & Eq (\ref{eq:gch})        & -0.2  &  $5\times 10^4$  &  $10^{15}$          &  -0.8  \\
2d$_i$   &  0.35  & 1.5   & Eq (\ref{eq:gch})        & -0.2  &  $5\times 10^4$  &  $10^{15}$          &  -0.8  \\
2b$_j$   &  0.35  & 1.0   & Eq (\ref{eq:gch})        &  0.1  &  $5\times 10^4$  &  $10^{15}$          &  -0.8  \\
2c$_j$   &  0.35  & 1.25  & Eq (\ref{eq:gch})        &  0.1  &  $5\times 10^4$  &  $10^{15}$          &  -0.8  \\
2d$_j$   &  0.35  & 1.5   & Eq (\ref{eq:gch})        &  0.1  &  $5\times 10^4$  &  $10^{15}$          &  -0.8  \\
2b$_k$   &  0.35  & 1.0   & Eq (\ref{eq:gch_broken}) &  0.5  &  $10^5$          &  $10^{15}$          &  -0.8  \\
2c$_k$   &  0.35  & 1.25  & Eq (\ref{eq:gch_broken}) &  0.5  &  $10^5$          &  $10^{15}$          &  -0.8  \\
2d$_k$   &  0.35  & 1.5   & Eq (\ref{eq:gch_broken}) &  0.5  &  $10^5$          &  $10^{15}$          &  -0.8  \\
2b$_l$   &  0.35  & 1.0   & Eq (\ref{eq:gch_broken}) &  1.0  &  $10^5$          &  $10^{15}$          &  -0.8  \\
2c$_l$   &  0.35  & 1.25  & Eq (\ref{eq:gch_broken}) &  1.0  &  $10^5$          &  $10^{15}$          &  -0.8  \\
2d$_l$   &  0.35  & 1.5   & Eq (\ref{eq:gch_broken}) &  1.0  &  $10^5$          &  $10^{15}$          &  -0.8  \\
2b$_m$   &  0.35  & 1.0   & Eq (\ref{eq:gch_broken}) &  1.0  &  $2\times 10^5$  &  $2\times 10^{15}$  &  -0.8  \\
2c$_m$   &  0.35  & 1.25  & Eq (\ref{eq:gch_broken}) &  1.0  &  $2\times 10^5$  &  $2\times 10^{15}$  &  -0.8  \\
2d$_m$   &  0.35  & 1.5   & Eq (\ref{eq:gch_broken}) &  1.0  &  $2\times 10^5$  &  $2\times 10^{15}$  &  -0.8  \\
2d$_m$2  &  0.35  & 1.5   & Eq (\ref{eq:gch_broken}) &  1.0  &  $2\times 10^5$  &  $2\times 10^{15}$  &  -0.7  \\
2d$_m$3  &  0.35  & 1.5   & Eq (\ref{eq:gch_broken}) &  1.0  &  $2\times 10^5$  &  $2\times 10^{15}$  &  -0.9
\enddata
\label{table:parameters}
\tablenotetext{a}{Names of the twenty numerical models presented in the paper;}
\tablenotetext{b}{Power-law index in the profile of the bulk Lorentz factor $\Gamma(r)$ in Equation (\ref{eq:bulk});}
\tablenotetext{c}{Power-law index in the profile of the magnetic-field strength $B(r)$ in Equation (\ref{eq:B});}
\tablenotetext{d}{Profile $\gch(r)$ for the characteristic Lorentz factor of electrons;}
\tablenotetext{e}{Single or broken power-law index in the $\gch$ profile in Equation (\ref{eq:gch}) or (\ref{eq:gch_broken});}
\tablenotetext{f}{Normalization value for the $\gch$ profile in Equation (\ref{eq:gch}) or (\ref{eq:gch_broken});}
\tablenotetext{g}{Normalization radius for the $\gch$ profile in Equation (\ref{eq:gch}) or (\ref{eq:gch_broken});}
\tablenotetext{h}{Low-energy photon index of the Band-function shape $H(x)$ in Equation (\ref{eq:Hx}).}
\end{deluxetable}
%--------------------------------------------------------

%--------------------------------------------------------
%
% Figures
%
%--------------------------------------------------------

\clearpage

%--------------------------------------------------------
%-------------------------------------------------------- 

%--------------------------------------------------------
%--------------------------------------------------------

%--------------------------------------------------------
%-------------------------------------------------------- 

\clearpage

%--------------------------------------------------------
%-------------------------------------------------------- 

\clearpage

%--------------------------------------------------------
%--------------------------------------------------------

\clearpage

%--------------------------------------------------------
%-------------------------------------------------------- 

\clearpage

%--------------------------------------------------------
%--------------------------------------------------------

\clearpage

%--------------------------------------------------------
%-------------------------------------------------------- 

\clearpage

%--------------------------------------------------------
%--------------------------------------------------------

\clearpage

%--------------------------------------------------------
%-------------------------------------------------------- 

\clearpage

%--------------------------------------------------------
%--------------------------------------------------------

\clearpage

%--------------------------------------------------------
%-------------------------------------------------------- 

\clearpage

%--------------------------------------------------------
%--------------------------------------------------------

\clearpage

%--------------------------------------------------------
%-------------------------------------------------------- 

\clearpage

%--------------------------------------------------------
%--------------------------------------------------------

\end{document}